\documentclass[apj]{emulateapj}
\usepackage{natbib}
\usepackage{graphicx}
\usepackage{txfonts}
\usepackage{amstext}
\usepackage{times}
\newcommand{\AAp}{\aap}
\newcommand{\ApJ}{\apj}

\newcommand{\CelMech}{Cel. Mech. Dyn. Astron.}

\newcommand{\be}{\begin{equation}}
\newcommand{\ee}{\end{equation}}
\newcommand{\bd}{\begin{displaymath}}
\newcommand{\ed}{\end{displaymath}}
\newcommand{\bea}{\begin{eqnarray}}
\newcommand{\eea}{\end{eqnarray}}
\newcommand{\sbs}[1]{\ensuremath{_\text{#1}}} 
\newcommand{\mum}{\ensuremath{\,\mu\m}}
\newcommand{\m}{\mathrm{m}}

\newcommand{\mm}{\mathrm{mm}}
\newcommand{\cm}{\mathrm{cm}}

\newcommand{\km}{\mathrm{km}}
\newcommand{\AU}{\mathrm{AU}}
\newcommand{\g}{\mathrm{g}}
\newcommand{\s}{\mathrm{s}}

\newcommand{\yr}{\mathrm{yr}}
\newcommand{\Myr}{\mathrm{Myr}}

\newcommand{\Jy}{\mathrm{Jy}}
\newcommand{\mJy}{\mathrm{mJy}}
\newcommand{\erg}{\mathrm{erg}}
\newcommand{\K}{\mathrm{K}}
\newcommand{\D}{\mathrm{d}}
\newcommand{\ain}{a_{\mathrm{inner}}}
\newcommand{\aout}{a_{\mathrm{outer}}}
\newcommand{\emax}{e_{\mathrm{max}}}
\newcommand{\imax}{i_{\mathrm{max}}}
\def\la{~\raise.4ex\hbox{$<$}\llap{\lower.6ex\hbox{$\sim$}}~}
\def\ga{~\raise.4ex\hbox{$>$}\llap{\lower.6ex\hbox{$\sim$}}~}
\submitted{Submitted to ApJ}
\received{2009 September 7}
\accepted{2009 November 30}
\shorttitle{The Debris Disk of Vega}
\shortauthors{M{\"u}ller et al.}

\begin{document}


\title{The Debris Disk of Vega: A Steady-State Collisional Cascade, Naturally}

\author{S. M{\"u}ller
       }
\affil{Astrophysikalisches Institut und Universit{\"a}tssternwarte, 
       Friedrich-Schiller-Universit\"at Jena,\\
       Schillerg\"a{\ss}chen~ 2--3, 07745 Jena, Germany
      }
\email{sebastian@astro.uni-jena.de}

\author{T. L{\"o}hne
       }
\affil{Astrophysikalisches Institut und Universit{\"a}tssternwarte, 
       Friedrich-Schiller-Universit\"at Jena,\\
       Schillerg\"a{\ss}chen~ 2--3, 07745 Jena, Germany
      }
\author{A. V. Krivov
       }
\affil{Astrophysikalisches Institut und Universit{\"a}tssternwarte, 
       Friedrich-Schiller-Universit\"at Jena,\\
       Schillerg\"a{\ss}chen~ 2--3, 07745 Jena, Germany
      }
\begin{abstract}
It has been argued that the photometric data and images of the
archetypical debris disk around Vega may be in contradiction with the
standard, steady-state collisional scenario of the disk evolution.
Here we perform physical modeling of the Vega disk ``from the sources''.
We assume that dust is maintained by a ``Kuiper belt'' of parent
planetesimals at $\sim 100$~AU and employ our collisional and radiative
transfer codes to consistently model the size and radial distribution of
the disk material and then thermal emission of dust.
In doing so, we vary a broad set of parameters, including the stellar
properties, the exact location, extension, and dynamical excitation of the
planetesimal belt, chemical composition of solids, and the collisional
prescription.
We are able to reproduce the spectral energy distribution in the entire
wavelength range from the near-infrared to millimeter, as well as the
mid-IR and sub-millimeter radial brightness profiles of the Vega disk.
Thus our results suggest that the Vega disk observations are compatible
with a steady-state collisional dust production, and we put important
constraints on the disk parameters and physical processes that sustain it.
The total disk mass in $\la 100~\km$-sized bodies is estimated to be
$\sim 10$ Earth masses.
Provided that collisional cascade has been operating over much of the Vega
age of $\sim 350~\Myr$, the disk must have lost a few Earth masses of
solids during that time.
We also demonstrate that using an intermediate luminosity of the star
between the pole and the equator, as derived from its fast rotation, is
required to reproduce the debris disk observations.
Finally, we show  that including cratering collisions into the model is
mandatory.
\end{abstract}
%
\section{INTRODUCTION}

Debris disks -- disks of planetesimals and dust around main-sequence
stars -- are known to encircle a sizeable fraction of main-sequence (MS)
stars.
The observed dust
is believed to derive from collisions between unobserved  planetesimals
that were neither used to make up planets nor ejected from the system by
the time when the nebular gas was dispersed
\citep[see][for a review]{Wyatt-2008}.

A standard scenario of a debris disk evolution
\citep[e.g.][]{thebault-et-al-2003,Krivov-et-al-2006,Strubbe-Chiang-2006,%
Thebault-Augereau-2007,Loehne-et-al-2008,Krivov-et-al-2008,Thebault-Wu-2008}
can be summarized as follows.
There is a relatively narrow belt of planetesimals (``birth ring'') in
orbits with moderate eccentricities and inclinations, exemplified by the
classical Kuiper belt in the solar system.
The orbiting planetesimals in the birth ring undergo collisional cascade
that grinds the solids down to dust.
At smallest dust sizes, stellar radiation pressure effectively reduces
the mass of the central star and quickly (on the dynamical timescale)
sends the grains into more eccentric orbits, with their pericenters still
residing within the birth ring while the apocenters are located outside
the ring.
As a result, the dust disk extends outward from the planetesimal belt.
The smaller the grains, the more extended their ``partial'' disk.
The tiniest dust grains, for which the radiation pressure effectively
reduces the physical mass by half, are blown out of the system in
hyperbolic orbits.
The radiation pressure blowout of the smallest collisional debris
represents the main mass loss channel in such a disk.
A steady state implies a balance between the production of dust by the
collisional cascade and its losses by radiation pressure blowout.

Such disks are usually referred to as collision-dominated, as opposed to
systems that~--- at dust sizes~--- are transport-dominated
\citep[e.g.][]{Krivov-et-al-2000}.
In the latter case, radial transport of dust material by various drag
forces occurs on shorter timescales then collisions.
Then, additional removal mechanisms may play a significant role.
For example, Poynting-Robertson (P-R) drag can bring grains close to the
star where they would sublimate or deliver them into the planetary region
where they would be scattered by planets.
To be transport-dominated, the system should either have an optical depth
below current detection limits \citep{Wyatt-2005} or be subject to
transport mechanisms other than P-R drag, such as strong stellar winds
around late-type stars \citep{Strubbe-Chiang-2006,Augereau-Beust-2006}.
Most of debris disks detected so far are thought to be
collision-dominated.
Accordingly, transport-dominated systems are not considered here.

The size segregation in collision-dominated disks described above means
that at different wavelengths the same disk would look differently
\citep[see, e.g., Fig. 17 in][]{Thebault-Augereau-2007}.
Measurements at longer wavelengths (sub-mm) probe larger grains, because
they are cooler, and thus trace the parent ring.
Such observations may also reveal clumps, if for instance there is a
planet just interior to the inner rim of the parent ring, and parent
planetesimals and their debris are trapped in external resonances,
similar to Plutinos in $3:2$ resonance with Neptune
\citep{Wyatt-2006,krivov-et-al-2006b}.
At shorter wavelengths (far-IR, mid-IR), smaller (warmer) grains are
probed.
Thus the same disk appears much larger. It may appear featureless, even
if the parent ring is clumpy, because strong radiation pressure
\citep{Wyatt-2006} and non-negligible relative velocities 
\citep{krivov-et-al-2006b} would liberate such particles from resonant
clumps.
As a result, they would form an extended disk, as described above,
regardless of whether their parent bodies are resonant or not.

The observed statistics of infrared (IR) luminosities of debris disks
around AFGK stars, including their dependence on the stellar age and
spectral class, is largely consistent with the steady-state collisional
evolution scenario presented above \citep{Wyatt-et-al-2007b,%
Loehne-et-al-2008}.
Furthermore, a detailed spectral energy distribution (SED) from far-IR to
mm wavelengths, which was measured for a handful stars, can be reproduced
with the models of a steady-state collisional cascade
\citep{Krivov-et-al-2008}.
There are some exceptions, however
\citep[see, e.g.,][and references therein]{Wyatt-2008}.
For instance, some A-type stars have fractional luminosities that are too
high compared to what is expected from collisionally evolving Kuiper belt
analogs.
A few percent of FGK stars show excess emission shortward of $24~\mum$
that must come from ``asteroidal'' region inside $\sim 10~\AU$, which in
some cases is also too high to be compatible with the standard scenario
\citep{Wyatt-et-al-2007}.
Besides, incidences and properties of debris disks around M-type stars
remain uncertain:
the observations are scarce, and the physics of such disks may be
different due to low stellar luminosity and strong stellar winds
\citep[e.g.,][]{Strubbe-Chiang-2006,Augereau-Beust-2006}.
Despite these caveats, the majority of the debris disks discovered so far
appears compatible with the steady-state collisional scenario.
However, the vast majority of debris disks are as yet unresolved, and the
set of observables is typically limited to the fractional luminosity and
a few photometric points in the IR.
Many more constraints on the disk properties could be posed by resolved
images, if these available, especially at more than one wavelength.
There are currently a dozen of disks with such datasets.
Accordingly, in this paper we undertake probably one of the first
attempts to investigate, whether all available data on one particular
debris disk star are compatible with a steady-state collisional scenario
of dust production and evolution.
We choose Vega.

Vega was the first MS star around which an IR excess over the
photospheric flux, indicative of debris dust, was discovered
\citep{Aumann-et-al-1984}.
It is therefore treated as an archetypical debris disk star;
MS stars with IR excesses are often  named ``Vega-type stars''.
Fluxes at wavelengths from mid-IR to millimeter have been measured with
{\it IRAS} \citep{Aumann-et-al-1984, Walker-Wolstencroft-1988}, {\it KAO}
\citep{Harper-et-al-1984,Harvey-et-al-1984}, and {\it ISO}
\citep{Heinrichsen-et-al-1998}.
As a result, its SED is known relatively
well.
The Vega disk has been resolved in sub-mm and mm with {\it JCMT}
\citep{Zuckerman-Becklin-1993, Holland-et-al-1998}, {\it OVRO}
\citep{Koerner-et-al-2001}, {\it IRAM} \citep{Chini-et-al-1990,%
Wilner-et-al-2002}, and {\it CSO} \citep{Marsh-et-al-2006}.
These observations reveal a clumpy ring of large dust grains between
about $80$ and $120~\AU$, suggesting a Kuiper belt analog.
\citet{Wyatt-2003} and \citet{Reche-et-al-2008} naturally explain the
ring structure with a resonant trapping of dust parent bodies by a
presumed Neptune- to Saturn-mass planet during their outward migration in
the past.

\citet{Su-et-al-2005} resolved the Vega system by {\it Spitzer/MIPS} at
24, 70, $160\mum$.
They found a featureless, huge disk extending up to $\sim 800~\AU$.
Although it came as a surprise at the time when this discovery was made,
it is no longer astonishing now.
As noted above, this is exactly what is expected:
a Kuiper belt-sized, clumpy ring of large dust grains seen in the sub-mm
and a much more extended disk of small grains, producing a smooth
brightness distribution evident in the mid- to far-IR.
However, by fitting these data, \citet{Su-et-al-2005} deduced
a mysterious overabundance of blowout grains of $\sim 1\mum$ in radius.
Under the assumption of a steady-state collisional disk evolution over
the Vega age, this would imply that the disk must have lost $\sim 3$
Jupiter masses of material, which appeared unlikely.
Accordingly, they suggested a recent major collisional event as a
possible explanation.
More exotic alternative scenarios proposed to explain such a large
fraction of blowout grains include a close stellar encounter
\citep{Makarov-et-al-2005} and a dynamical instability event similar to
what caused the Late Heavy Bombardment  in the solar system
\citep{Wyatt-et-al-2007}.
However, \citet{Kenyon-Bromley-2008}, who modeled the Vega debris disk as
an aftermath of icy planet formation with their hybrid multi-annulus
coagulation code, find their model to be capable of reproducing the
{\it Spitzer} fluxes, questioning the need in alternative scenarios.

We note that the excessive amount of grains in blowout orbits inferred
by \citet{Su-et-al-2005} uncovers another problem.
A steady-state collisional evolution implies a certain size distribution
of dust.
Typically, the amount of blowout grains instantaneously present in the
steady-state system is much less than the amount of slightly larger
grains in loosely bound orbits around the star.
This is because the dust production of the grains of adjacent sizes is
comparable, but the lifetime of bound grains (due to collisions) is much
longer than  the lifetime of blowout grains (disk-crossing timescale).
A jump in the size distribution around the blowout size is a robust
prediction of all collisional cascade models, even those that do not
assume a steady state (e.g. those that describe short-lived consequences
of a major break-up event).
Thus the amount of blowout grains reported by \citet{Su-et-al-2005} would
necessitate an unrealistically huge amount of larger grains in bound
orbits.
And this conclusion would hold not only in a steady-state scenario, but
also all alternative scenarios listed above would face the same problem.

Beside the outer disk, the inner part of the system reveals another
peculiarity.
Pioneering interferometry observations with {\it CHARA/FLUOR} in the
near-IR \citep{Absil-et-al-2006} have led to the discovery of a dust
cloud just exterior of the sublimation zone, well inside $1~\AU$
(``exozodi'').

Just like the dust in the system, the central star turned out to be
unusual, too.
\citet{Peterson-et-al-2006} and \citet{Aufdenberg-et-al-2006} found Vega
to be a rapid rotator, which makes stellar parameters functions of the
stellar latitude.
Table~\ref{tab:Vega_Par} summarizes the stellar parameters relevant for
this study.
It remains unclear whether unusual properties of the disk are somehow
related to those of the star.

\begin{table}[t!]
  \caption{Stellar parameters.}
  \label{tab:Vega_Par}
  \begin{center}
  \renewcommand{\arraystretch}{1.1}
  \tabcolsep 3pt
  \begin{tabular}{lccc}
    \tableline\tableline
                              & Equator                 & Pole            & Note \\
    \tableline
    $R~[R_\odot]$             & $2.873\pm0.026$         & $2.306\pm0.031$ & 1    \\
    $T_{\mathrm{eff}}~[\K]$   & $7\,900^{+500}_{-400}$  & $10\,150\pm100$ & 2    \\
    $L_*~[L_\odot]$           & $28^{+8}_{-6}$          & $57\pm3$        & 3    \\ 
    $\log (g[\cm \s^{-2}])$   & $4.074\pm0.012$         & $3.589\pm0.056$ & 1    \\
    \hline
    $M_*~[M_\odot]$           & \multicolumn{2}{c}{$2.3\pm0.2$}           & 2    \\
    $\mathrm{Age}~[\Myr]$     & \multicolumn{2}{c}{$350$}                 &      \\
    \tableline
  \end{tabular}
  \end{center}
{\small
           Notes:
           (1) From \citet{Peterson-et-al-2006},
           (2) From \citet{Aufdenberg-et-al-2006},
           (3) Luminosity at the equator and at the poles derived from the equatorial
               and polar values of the stellar radius $R$ and effective temperature $T_{\mathrm{eff}}$
               and from the average stellar luminosity of $37 \pm 3~L_\odot$ 
               \citep{Aufdenberg-et-al-2006}
               through the Stefan-Boltzmann relation.
	}
\end{table}

In this paper, we re-address the question of whether the available data
are compatible with a steady-state collisional scenario of dust
production and evolution.
Instead of simply seeking dust distributions -- e.g. in the form of power
laws -- that would provide the best fit to the observables, we employ an
approach described by \citet{Krivov-et-al-2008}.
In this approach, we assume a planetesimal belt with certain properties,
evolve it with a collisional code to generate the dust portion of the
debris disk, calculate the emission of that dust, and compare it to the
observed emission.
The best fit can be achieved by varying the parameters that describe the
planetesimal belt (e.g. its location and mass), rather than parameters of
the dust distribution as is commonly done.

In Sect. \ref{sec:datareduction} we describe the data and their reduction
and in Sect. \ref{sec:methods} dynamical and thermal emission models used
in this paper.
Section \ref{sec:reference} presents our reference (``first-guess'') disk
model and compares it with the available observational data.
In Sect. \ref{sec:improvements} we analyze the influence of various model
parameters on the observables.
Section ~\ref{sec:discussion} contains a discussion and
Sect.~\ref{sec:summary} lists our conclusions.

\section{OBSERVATIONAL DATA} \label{sec:datareduction}

From the {\it Spitzer} archive, we extracted {\it MIPS} images of the
Vega system at 24, 70, and $160\mum$, using the {\it Leopard}
software\footnote{{\sf http://ssc.spitzer.caltech.edu/mips/}}.
After applying standard basic corrections, the data were further
processed to remove remaining constant backgrounds.
Column averages were used to eliminate the so-called jail-bar
artifacts caused by saturated sources on $24\mum$ images.

For each image, the center of the source was then determined by
minimizing first order moments.
For the $160\mum$ images, the pointing information was used for that
purpose.
After that, radial profiles were derived in an iterative process with two
steps:
(i) sampling the interpolated image at discretely binned distances around
the source center with a sub-pixel resolution and
(ii) integrating and subtracting the resulting profile pixel-wise from
the original.
Five such iterations were performed for each image.
Finally, all profiles for each wavelength were combined, using their
median.

\begin{figure}
  \centering
  \includegraphics[width=0.8\linewidth]{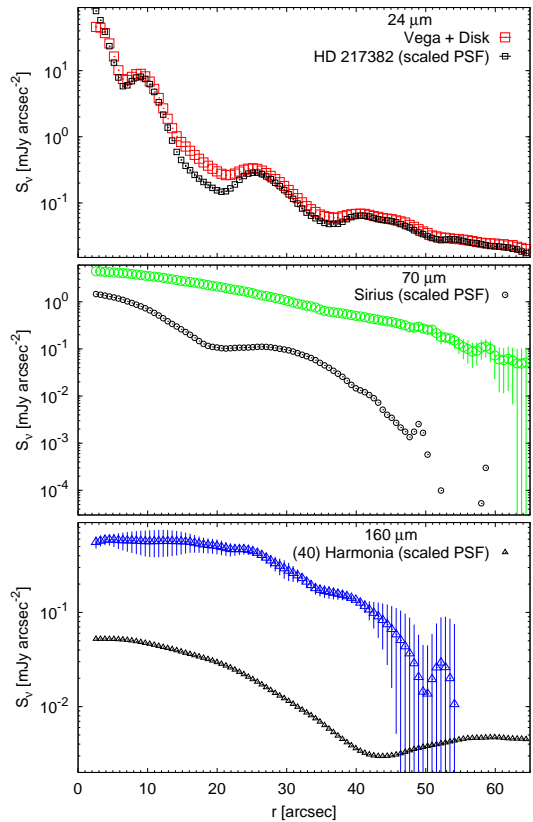}
  \caption{Radial profiles of the surface brightness for Vega as extracted
           from the {\it Spitzer MIPS} images for 24 (large squares), 70 (large  
           circles), and $160\mum$ (large triangles) from top to bottom.
           The standard sources
           HD~217382, Sirius, and (40) Harmonia used for PSF subtraction for
           the different wavelengths are shown as small symbols of the same 
           shapes.
          }
  \label{fig:psf}
\end{figure}

At $24\mum$, the data reduction is complicated by the fact that the
central part of the images ($< 4''$ or $<30$~AU) is saturated.
Still, after subtracting the reference Point Spread Function (PSF), the
outer part of the brightness profile can be found with sufficient
accuracy.
Figure~\ref{fig:psf} (top) shows the median profile extracted for Vega
compared to that of the reference star HD~217382, a K4III giant at a
distance of 120~AU, scaled up from its intrinsic 2.15~Jy to the 7.4~Jy
\citep{Su-et-al-2005} of Vega's photosphere.
The resulting photometric excess, obtained by integrating the
subtracted profile (the Vega profile minus the scaled PSF) from $4''$
outward, is $0.94 \pm 0.28~\Jy$.
Here the formal error is determined by the standard deviations of Vega
and the PSF plus an assumed systematic error of 5~\% in the photometry of
Vega's photosphere.
However, the flux is probably more uncertain because so is the
contribution of the inner part of the disk, between $4''$ and $10''$
($30$--$80$~AU).
Thus, for a more accurate comparison with our models below, we can use
the ``certain'' part of the observed flux by integrating the brightness
profile from $10''$ outward.
This nearly halves the total $24\mum$ flux, giving $0.53~\Jy$.
This ``partial'' flux will later be compared with the flux predicted by
our models exactly in the same range of distances, from $10''$ ($80$~AU)
outward.
Coincidentally, it is this range where the emitting dust is only present
in most of our models, because the inner edge of the birth ring is
located at $80$~AU.
We emphasize, however, that the true total $24\mum$ excess flux may be
higher.
It probably lies in the range $0.5$--$0.9~\Jy$, with no obvious
possibility to narrow it because of the saturation problem described
above.

The reduction of the $70\mum$ images was more straightforward, resulting
in the radial profiles shown in Fig.~\ref{fig:psf} (middle) and an excess
of $8.6\pm 1.2$~Jy. Sirius (HD 48915) was used as reference.

Since stray light strongly contaminates images of hot, i.e.~stellar,
sources at $160\mum$, we followed the strategy of
\citet{Stapelfeldt-et-al-2004} and used an additional reference star
(HD 197989) to remove this artifact.
However, visual inspection of the Vega images already suggested that the
removal procedure would not be perfect because of some dissimilarities in
the artifact shapes.
Therefore, only the half-image facing away from the artifact center was
used for profile extraction, as was done by \citet{Su-et-al-2005}.
The actual reference used for PSF subtraction was the asteroid (40)
Harmonia, although its influence on the excess of $1.9_{-0.5}^{+0.8}$~Jy
is negligible.
See Fig.~\ref{fig:psf} (bottom) for the resulting profiles.

The photometry points across the wavelength range from mid-IR to
millimeter taken from the literature, are listed in 
Table~\ref{tab:photometry}.
These points are plotted as symbols with error bars in bottom left panels
of Figs. \ref{fig:time-mod}--\ref{fig:best-fit}.

\begin{table}[t!]
  \caption{Photometric data.}
  \label{tab:photometry}
  \begin{center}
  \scriptsize
  \tabcolsep 3pt
  \begin{tabular}{rlll}
    \tableline\tableline
    $\lambda~[\mum]$ & $F_{disk}~[\mJy]$               & Instr.	    & Ref.  \\
    \tableline
     25.0            & $(2.979 \pm 0.936) \times 10^3$ & IRAS       & \citet{Walker-Wolstencroft-1988}\\
     25.0            & $(4.019 \pm 2.080) \times 10^3$ & ISO        & \citet{Heinrichsen-et-al-1998}\\
     47.0            & $(6.414 \pm 1.640) \times 10^3$ & KAO        & \citet{Harvey-et-al-1984}\\
     60.0            & $(7.918 \pm 0.901) \times 10^3$ & IRAS       & \citet{Walker-Wolstencroft-1988}\\
     60.0            & $(9.318 \pm 2.082) \times 10^3$ & ISO        & \citet{Heinrichsen-et-al-1998}\\
     80.0            & $(9.091 \pm 2.910) \times 10^3$ & ISO        & \citet{Heinrichsen-et-al-1998}\\
     95.0            & $(6.829 \pm 1.825) \times 10^3$ & KAO        & \citet{Harvey-et-al-1984}\\
    100.0            & $(7.109 \pm 0.754) \times 10^3$ & IRAS       & \citet{Walker-Wolstencroft-1988}\\
    100.0            & $(5.969 \pm 1.920) \times 10^3$ & ISO        & \citet{Heinrichsen-et-al-1998}\\
    120.0            & $(4.890 \pm 1.551) \times 10^3$ & ISO        & \citet{Heinrichsen-et-al-1998}\\
    170.0            & $(2.437 \pm 0.771) \times 10^3$ & ISO        & \citet{Heinrichsen-et-al-1998}\\
    193.0            & $(8.932 \pm 5.000) \times 10^2$ & KAO        & \citet{Harper-et-al-1984}\\
    200.0            & $(1.321 \pm 0.408) \times 10^3$ & ISO        & \citet{Heinrichsen-et-al-1998}\\
    350.0            & $(4.691 \pm 1.500) \times 10^2$ & CSO        & \citet{Marsh-et-al-2006}\\
    450.0            & $(1.301 \pm 0.450) \times 10^2$ & CSO        & \citet{Marsh-et-al-2006}\\
    800.0            & $(1.626 \pm 0.500) \times 10^1$ & IRAM       & \citet{Chini-et-al-1990}\\
    800.0            & $(2.426 \pm 1.500) \times 10^1$ & JCMT       & \citet{Zuckerman-Becklin-1993}\\
    850.0            & $(4.086 \pm 0.540) \times 10^1$ & JCMT       & \citet{Holland-et-al-1998}\\
    870.0            & $(1.721 \pm 0.900) \times 10^1$ & IRAM       & \citet{Chini-et-al-1990}\\
   1300.0            & $(2.390 \pm 1.500) \times 10^0$ & IRAM       & \citet{Chini-et-al-1990}\\
   1300.0            & $(1.189 \pm 0.190) \times 10^1$ & OVRO       & \citet{Koerner-et-al-2001}\\
   1300.0            & $(1.140 \pm 0.170) \times 10^1$ & IRAM       & \citet{Wilner-et-al-2002}\\
    \tableline
  \end{tabular}
  \end{center}
\end{table}

\section{METHODS} \label{sec:methods}

\subsection{Dynamical and Collisional Evolution} \label{sec:ace}

Our technique to follow the size and radial distribution of material in
rotationally-symmetric debris disks is described in detail in previous
papers \citep{Krivov-et-al-2000,Krivov-et-al-2005,Krivov-et-al-2006,%
Krivov-et-al-2008,Loehne-et-al-2008}.
Our numerical code,  {\it ACE} ({\it Analysis of Collisional Evolution}),
solves the Boltzmann-Smoluchowski kinetic equation over a grid of masses,
periastron distances, and orbital eccentricities of solids.
It includes the effects of stellar gravity, direct radiation pressure,
drag forces, as well as disruptive and erosive collisions.

The results of {\it ACE} simulations depend sensitively on the adopted
collisional  prescription.
Possible collisional outcomes range from a perfect sticking or
fragmentation with subsequent reaccumulation of particles (when the
impact energy is low) to a crater formation or even a complete disruption
(when it is high).
An important quantity in the collisional prescription is the critical
specific energy for disruption and dispersal, $Q\sbs{D}^\star$.
It is defined as the impact energy per unit target mass that results in
the largest remnant containing a half of the original target mass.
For small objects, $Q\sbs{D}^\star$ is determined solely by the material
strength, while for objects larger than $\sim 100\,\m$, the gravitational
binding energy dominates.
As a result, $Q\sbs{D}^\star$ is commonly described by the sum of two
power laws \citep[see, e.g.,][]{Davis-et-al-1985,Holsapple-1994,%
Paolicchi-et-al-1996,Durda-Dermott-1997,Durda-et-al-1998,%
Benz-Asphaug-1999,Kenyon-Bromley-2004b}:
\begin{equation}
  Q\sbs{D}^\star 
   = Q\sbs{D,s}(s)
     \left(\frac{s}{1~\mathrm{m}}\right)^{-3b\sbs{s}}
   + Q\sbs{D,g}(s)
     \left(\frac{s}{1~\mathrm{km}}\right)^{3b\sbs{g}},
\label{eq:QD}
\end{equation}
where the subscripts s and g denote the strength and gravity regime,
respectively.

The thermal emission in the mid- and far-IR is dominated by small dust
with sizes slightly above the so-called blowout limit, below which
particles are repelled from the system by stellar radiation pressure.
An equilibrium size distribution set by the collisional cascade,
especially near that blowout limit, depends on the total mass and the
size distribution of fragments of each individual collision.
Depending on the ratio of the impact energy to the critical energy, the
average size of the fragments changes.
This has been measured \citep[e.g.][]{Takagi-et-al-1984} and numerically
modeled \citep[e.g.][]{Durda-et-al-2007}, and useful analytic
prescriptions for use in collisional codes are available 
\citep[e.g.][]{Thebault-Augereau-2007}.
Here we apply a simplified approach with only one parameter, the slope
$\eta$ of a single power law
$\mathrm{d} N \propto m^{-\eta} \mathrm{d} m$.
Larger values of $\eta$ put more weight on small fragments, while smaller
values of $\eta$ ``prefer'' larger fragments.
Note that using a single power law down to infinitely small grains is
only meaningful for $\eta < 2$.

\begin{figure}
  \centering
  \includegraphics[width=0.8\linewidth]{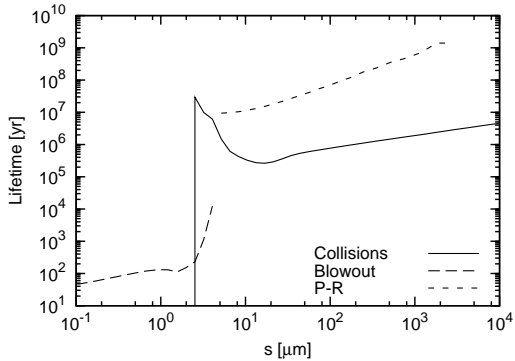}
  \caption{Timescales as functions of the grain size:
           collisional time (solid line),
           blowout time (long-dashed), and
           P-R time (short-dashed). 
           The collisional time is an average over the grain orbits
           with all possible pericentric distances $q$ and eccentricities $e$,
           weighted with the amounts of those particles in the disk.
           The P-R time is the time it takes for a grain to drift
           across the parent ring or, more exactly, the time interval
           over which a grain's $q$ reduces from 120~AU to 80~AU.
           It was computed by simultaneously solving the orbit-averaged 
           equations that describe $q(t)$ and $e(t)$
           \citep{Burns-et-al-1979}.         
           All timescales are for the parameters of the reference model 
           (see Sect.~\ref{sec:reference}).
          }
  \label{fig:timescales}
\end{figure}

Each {\it  ACE} run outputs, among other quantities, the size and radial
distribution of disk solids over a broad size range from sub-micrometers
to hundreds of kilometers at different time steps, and the code is fast
enough to evolve the distribution over gigayears.
As shown in Fig.~\ref{fig:timescales}, typical timescales for P-R drag in
the Vega disk are much longer than collisional lifetimes, except
for a very narrow size range close to the blowout limit, where
both become comparable.
Thus we switch off the P-R effect in the {\it ACE} runs, but
make additional checks in Sect.~\ref{sec:PR}.
Gas drag can safely be neglected, because the Vega system with 
its 350~Myr age could not have retained any primordial gas, while
the density of secondary gas cannot be high enough to influence the 
dust dynamics.
As long as the drag forces are discarded, the number of parameters to
vary, and thus the number of required {\it ACE} runs, can be reduced by
applying scaling laws derived in \citet{Loehne-et-al-2008} and
\citet{Krivov-et-al-2008}.
Specifically, it is not needed to perform separate {\it ACE} runs for the
initial planetesimal disk with different initial masses.

\subsection{Thermal Emission of Dust} \label{sec:emission}

Considering a rotationally symmetric disk of spherical particles, we
denote by $N(r,s)$ the disk's surface number density of grains of radius
$s$ at a distance $r$ from the star.
The specific thermal emission flux from the entire disk
$F_{\lambda, \mathrm{disk}}^{\mathrm{tot}}$, measured by an observer at a
distance $D$ from the star at a given wavelength $\lambda$, can be
calculated as \citep{Krivov-et-al-2008} 
\bea
  F_{\lambda, \mathrm{disk}}^{\mathrm{tot}}
            & = & \frac{2\pi^2}{D^2} \int r(T_g) \, \frac{\D r(T_g)}{\D T_g} \D T_g \int \D s \; s^2 \;\times\nonumber\\
            &   & \times \; N(r,s) \,Q_{\lambda}^{abs}(s) \, B_{\lambda}(T_g)
  \label{equ:F} ,
\eea
where $B_{\lambda}(T_{\mathrm{g}})$ is the Planck function,
$Q_{\lambda}^{\mathrm{abs}}(s)$ is the absorption efficiency, $\lambda$
is the wavelength, and the grain temperatures $T_{\mathrm{g}}(s)$ are
deduced from the standard assumption of thermal equilibrium with the
stellar radiation field.
Similarly, the radial profile of the surface brightness $S_\lambda(r)$ is
given by
\be
  S_\lambda(r) \approx \frac{\pi}{D^2} \int \D s \; s^2 N(r, s) Q^{\mathrm{abs}}_\lambda(s) B_\lambda(T_{\mathrm{g}}(r, s))
  \label{equ:SB} .
\ee
To enable comparison with observations, equation (\ref{equ:SB}) has to be
convolved with the normalized PSF of the
instrument:
\be
  S_\lambda^{\prime}(\mathbf{r}) = S_\lambda(\mathbf{r}) \otimes \mathrm{PSF}_\lambda(\mathbf{r})
  \label{equ:conv} .
\ee

A numerical solution of Eqs. (\ref{equ:F}) and (\ref{equ:SB}) has been
implemented in two routines, {\it SEDUCE} ({\it SED Utility for
Circumstellar Environment}) and {\it SUBITO} ({\it SUrface Brightness
Investigation TOol}), respectively.
The stellar flux needed to compute the dust temperatures is extracted
from {\it NextGen} models \citep{Hauschildt-et-al-1999}.
Both codes have a direct interface to {\it ACE}, so that {\it ACE} output
can be used as input to {\it SEDUCE} and {\it SUBITO}.
To arrive at the correct total dust mass and absolute values of fluxes,
we rescale the {\it ACE}--{\it SEDUCE} and {\it ACE}--{\it SUBITO}
simulation results with the aid of the appropriate scaling laws, as
described in Appendix~A of \citet{Krivov-et-al-2008}.

\section{THE REFERENCE MODEL OF THE VEGA DISK} \label{sec:reference}

\subsection{Choice of Model Parameters}

In the reference model, we use the stellar luminosity at the equator of
$L_\star = 28~L_\odot$ (Table~\ref{tab:Vega_Par}), and we assume that the
collisional cascade has been operating over the entire stellar age,
350~Myr.

According to \citet{Dent-et-al-2000}, \citet{Su-et-al-2005},
\citet{Marsh-et-al-2006} and \citet{Wyatt-2006} we adopt~--- as a
``first guess''~--- an initial ring of parent bodies with semimajor axes
ranging from 80 to 120~AU and an initially constant surface density
in this range.
The clumpy shape of the sub-mm ring, usually interpreted through resonant
capture of planetesimals by an unseen planet interior to the ring,
implies that the eccentricities of the planetesimals are not very low
\citep{Wyatt-2003,Reche-et-al-2008}.
On the other hand, the relatively narrow ring observed at wavelengths
longer than $350\mum$ indicates that eccentricities cannot be too high.
As a reasonable compromise and for the sake of simplicity, for the
reference model we adopt a uniform distribution of eccentricities from
0.0 to a moderate value of 0.2. 
Maximum orbital inclinations (or a semi-opening angle) were then set to
$0.1$~rad in accord with the energy equipartition assumption.
Thus, the initial planetesimal disk resides between 64 and 144~AU from
the star.
This is still in agreement with the observed 80 to 120~AU as most of the
material is concentrated in the central part of the initial ring.
Note that these assumptions describe the {\em initial} disk extension.
The subsequent collisional and dynamical evolution of the parent belt
slightly changes the distributions of planetesimals.

\begin{figure*}
  \centering
  \includegraphics[width=0.8\linewidth]{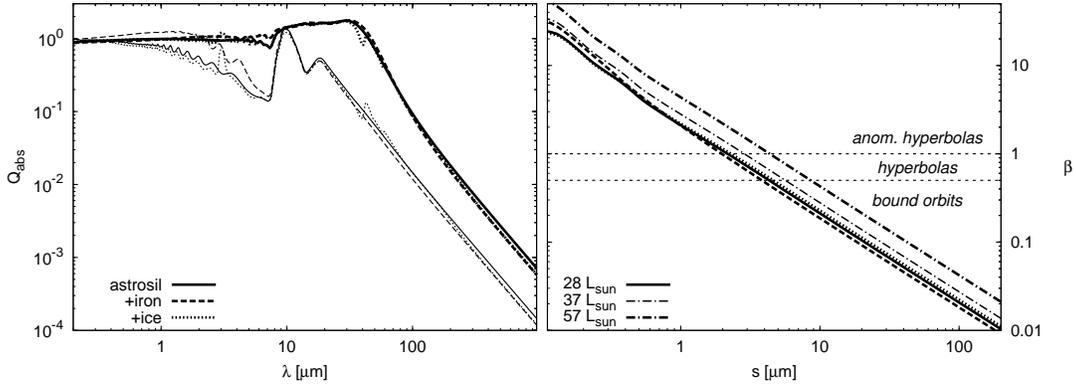}
  \caption{Left: Absorption efficiency for 1 (thin lines) and $4.9\mum$
                 (thick lines) particles consisting of astronomical
                 silicate (solid lines) and of an astronomic silicate
                 matrix with 10~\% iron (dashed lines) and water ice
                 (dotted lines) inclusions.
           Right: The $\beta$ ratio of the same grains as in the left panel 
                  (the same linestyles), assuming $L=28~L_\odot$.
                  Additionally, thin and thick dash-dotted lines show the $\beta$ ratio
                  for pure astrosilicate grains, but higher stellar luminosities of 
                  $L = 37~L_\odot$ and $57~L_\odot$ respectively.
          }
  \label{fig:beta}
\end{figure*}

All particles were assumed to be composed of astronomical silicate
\citep[$\rho = 3.3~\g\cm^{-3}$,][]{Laor-Draine-1993}.
Mie theory was used to calculate radiation pressure and absorption
efficiencies
(Fig.~\ref{fig:beta}).

The disk was modeled with {\it ACE} with grains ranging from $0.05\mum$
to $67~\km$ in radius and the mass ratio in the adjacent bins of $4$.
The pericenter distance grid covered 50 logarithmically-spaced values
from 20~AU to 800~AU.
The eccentricity grid contained 100 linearly-spaced values between
$-5.0$ and $5.0$ (eccentricities are negative in the case of smallest
grains with $\beta >1$, whose orbits are anomalous hyperbolas, open
outward from the star).
The distance grid used by {\it ACE} to output distance-dependent
quantities such as the size distribution was 10~AU through 600~AU at
10~AU increments.
In the collisional prescription, we set
$Q\sbs{D,s}(s) = Q\sbs{D,g}(s) = 5 \times 10^6~\erg\g^{-1}$,
$3 b\sbs{g} =1.38$, and $3 b\sbs{s} =0.37$
and take the size distribution of fragments to be a power law with index
$\eta = 1.833$.
Both disruptive and cratering collisions are switched on.

A set of parameters used for the reference model is given in the first
line of Tab. \ref{tab:parameters}.
In the table, we only list those parameters that we later vary with
respect to the reference model.

\begin{table*}[t!]
  \caption{Sets of model parameters used in the simulations.
           The parameters of the reference model are listed in the first line.
           For all other models, only parameters that are different from those of the
           reference model are given.
           }
  \label{tab:parameters}
  \begin{center}
  \footnotesize
  \tabcolsep 3pt
  \begin{tabular}{cccccccccccc}
    \tableline\tableline
    Model & $L_*~[L_\odot]$ & $T$~[Myr] & $\ain~[\AU]$      & $\aout~[\AU]$ & $\emax$       & $\imax$ & composition      & collisions              & $Q_{D,s}~[\erg\g^{-1}]$     & $b_s$     & $\eta$      \\ \tableline
    ref.  & 28              & 350       & 80                & 120               & 0.2       & 0.1            & no incl.         & w/ cratering            & $5.0\times 10^6$      & 0.37      & 1.833       \\ \tableline\\[-6pt]
    a1    & ---             &      35   & ---               & ---               & ---       & ---            & ---              & ---                     & ---                   & ---       & ---         \\
    a2    & ---             &      3.5  & ---               & ---               & ---       & ---            & ---              & ---                     & ---                   & ---       & ---         \\[4pt]
    b1    & ---             & ---       &      50           & ---               & ---       & ---            & ---              & ---                     & ---                   & ---       & ---         \\
    b2    & ---             & ---       &      100          & ---               & ---       & ---            & ---              & ---                     & ---                   & ---       & ---         \\
    b3    & ---             & ---       & ---               &      100          & ---       & ---            & ---              & ---                     & ---                   & ---       & ---         \\
    b4    & ---             & ---       & ---               &      150          & ---       & ---            & ---              & ---                     & ---                   & ---       & ---         \\[4pt]
    c1    &      37         & ---       & ---               & ---               & ---       & ---            & ---              & ---                     & ---                   & ---       & ---         \\
    c2    &      57         & ---       & ---               & ---               & ---       & ---            & ---              & ---                     & ---                   & ---       & ---         \\
    c3    & ---             & ---       & ---               & ---               & ---       & ---            &      ice incl.   & ---                     & ---                   & ---       & ---         \\
    c4    & ---             & ---       & ---               & ---               & ---       & ---            &      iron incl.  & ---                     & ---                   & ---       & ---         \\[4pt]
    d1    & ---             & ---       &      71.1         &      130.9        &      0.1  &      0.05      & ---              & ---                     & ---                   & ---       & ---         \\
    d2    & ---             & ---       &      91.4         &      100.8        &      0.3  &      0.15      & ---              & ---                     & ---                   & ---       & ---         \\[4pt]
    e1    & ---             & ---       & ---               & ---               & ---       & ---            & ---              &      w/o cratering      & ---                   & ---       & ---         \\
    e2    & ---             & ---       & ---               & ---               & ---       & ---            & ---              & ---                     & $    2.5\times 10^6$  &      0.2  & ---         \\
    e3    & ---             & ---       & ---               & ---               & ---       & ---            & ---              & ---                     & $    6.9\times 10^6 $ &      0.45 & ---         \\[4pt]
    f1    & ---             & ---       & ---               & ---               & ---       & ---            & ---              & ---                     & ---                   & ---       &      1.6    \\
    f2    & ---             & ---       & ---               & ---               & ---       & ---            & ---              & ---                     & ---                   & ---       &      1.95   \\[4pt] \tableline
    fit   &      45         & ---       &      62           &      120          &      0.1  &      0.05      & ---              & ---                     & ---                   & ---       &      1.95   \\ \tableline
  \end{tabular}
  \end{center}
\end{table*}

\subsection{Size, Radial, and Temperature Distributions}

\begin{figure*}
  \centering
  \includegraphics[width=0.7\linewidth]{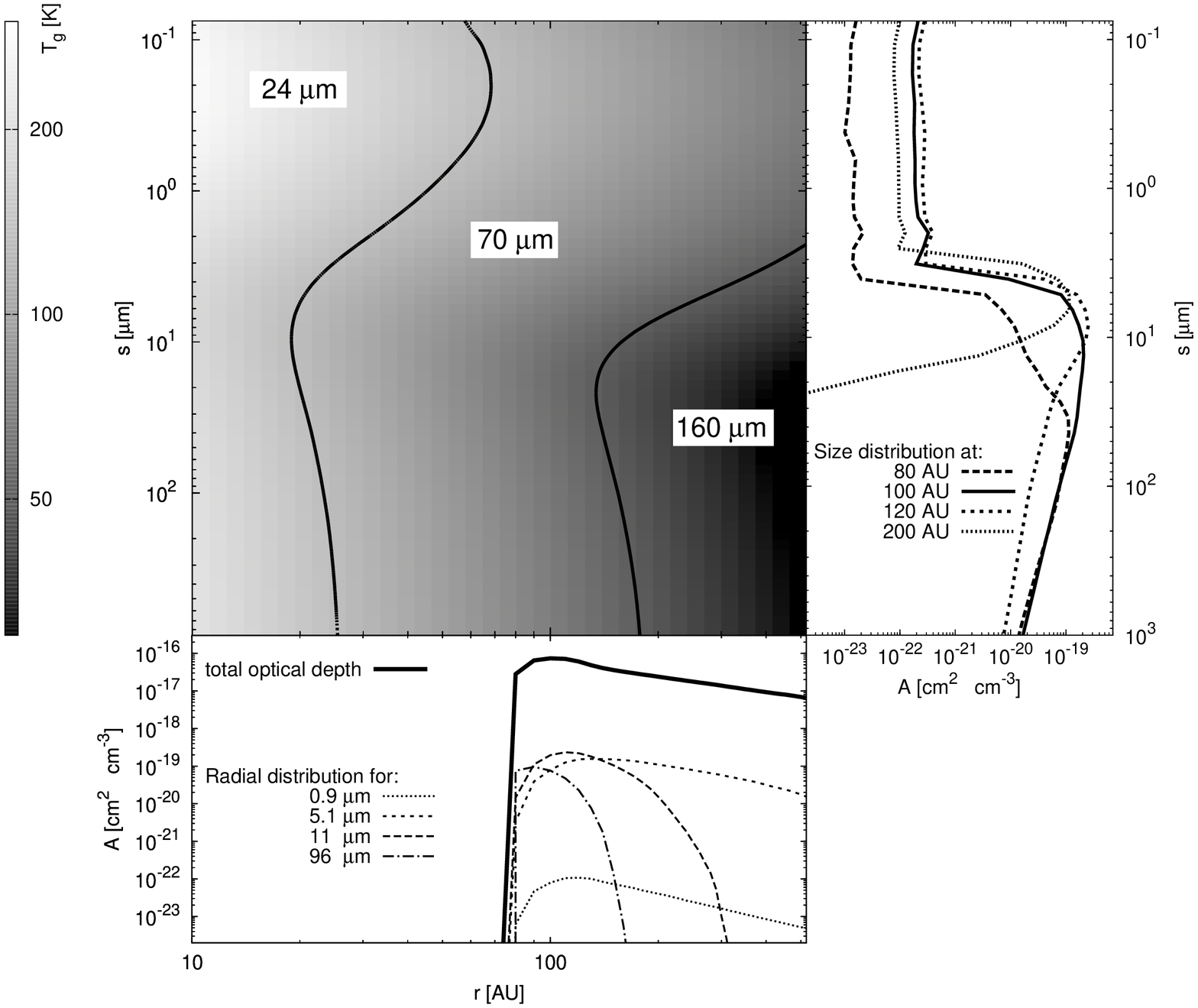}
  \caption{Left top: Disk's temperature profile for the assumed astrosilicate
                     grains and stellar parameters as derived for Vega's
                     equator. Two solid lines separate the regions of
                     dominant emission at the three {\it Spitzer/MIPS}
                     wavelengths: in the left part $24\mum$ emission is
                     more efficient than emission at the other two
                     wavelengths, in the central part $70\mum$
                     emission dominates, and in the right part $160\mum$ emission is
                     the strongest.
           Left bottom: Radial distribution in the reference model
                        for 0.9, 5.1, 11 and $96\mum$ grains (thin lines)
                        and the resulting total optical depth in arbitrary units 
                        (thick solid line).
           Right: Grain size distribution in the reference disk
                  at 80, 100, 120, and 200~AU.
          }
  \label{fig:T+dist}
\end{figure*}

Dust distributions in our reference model are presented in
Fig.~\ref{fig:T+dist}.
The right panel shows the grain size distributions within and outside the
birth ring.
The radiation pressure blowout effect causes a steep drop between 3 and
$5\mum$, which corresponds to $\beta \approx 0.5$ (Fig.~\ref{fig:beta}
right).
As previous studies have shown, the blowout drop in the size distribution
results in a more or less pronounced wavy pattern in the distribution of
larger particles \citep[e.g.][]{Thebault-et-al-2003,Krivov-et-al-2006},%
with the ``wavelength'' and amplitude of the pattern depending on
material strength and impact velocities.
However, compared to previous studies \citep[e.g.][]{Krivov-et-al-2006},
{\it ACE} now uses a grid of pericentric distances instead of semimajor
axes.
This reduces the effects of discretization on the effective relative
velocities, which is strongest for particles on highly eccentric orbits
with pericenters close to the birth ring.
Therefore, the waviness is washed out, especially in the reference run.

The radial distribution of different-sized grains is shown in the bottom
panel of Fig. \ref{fig:T+dist} with thin lines.
As explained above, most of the material is located between 80 and
120~AU\footnote{Note that the ring starts at about 73~AU and not at the
above-mentioned 64~AU.
This is due to the eccentricity binning.
Individual bins are 0.1 wide and centered at 0.05, 0.15, and so on.
Thus, the largest effective eccentricity for
$e\sbs{max} = 0.2$ is $e = 0.15$.
The corresponding  minimum pericentric distance for $a = 80~AU$ is
$q\sbs{min} = 0.85 \times 80~\AU = 68\AU$.
The next larger point in the pericenter grid is then centered at 73~AU.}.
The largest particles are confined to this region as they are nearly
unaffected by radiation pressure.
The smaller the particles, the wider they are spread over the disk.
In addition, the bottom panel in Fig. \ref{fig:T+dist} plots the total
normal geometrical optical depth $\tau$ (in arbitrary scale).
The optical depth beyond about 120~AU is dominated by particles just
above the blowout size, $\sim 5\mum$, which are in barely bound orbits.
It is only in the region of the birth ring where particles with
$s \ga 10\mum$ make a significant contribution to the optical depth.

To judge about the dust temperatures and the thermal emission of the disk
in the reference model, the left top panel in Fig. \ref{fig:T+dist}
depicts the dust 
temperature as a function of stellar distance and grain size.
Both distance and size axes share those of the two other panels.
Hence, extrapolating the maxima of the size and radial distributions into
the temperature plot yields a ``typical'' temperature, i.e. the
temperature of grains with $\tau$-dominating sizes at the distance where
$\tau$ peaks.
For our reference disk these are grains with $s \approx 5$...$10\mum$
at $r \sim 80$...$120$~AU and their temperature is about 60~K.

\subsection{Dust Mass and Disk Mass}

We then used {\it SEDUCE} to obtain the SED or the reference disk and to
fit it vertically to the available observations starting at $25\mum$.
Shorter wavelengths were neglected in the fitting process as the
uncertainty of the photospheric subtraction there is too high.
Furthermore, we did not use the {\it Spitzer} $24\mum$ data point for
fitting because of the uncertainties in converting the images into
photometry, as discussed in Sect.~\ref{sec:datareduction}).
In the SED calculation, we adopted the stellar parameters at the equator
to obtain the photosphere seen by the dust disk, but took the polar
values to calculate the observed photosphere (Tab. \ref{tab:Vega_Par}).

The aforementioned fitting itself is not as straightforward as it may
seem.
As we wish the modeled absolute thermal emission flux to match the
actually observed one, we need to change the amount of dust, i.e. the
dust mass.
In our approach, however, only the parameters of the parent planetesimals
can be changed, not those of dust they produce.
It means that we have to go one step back and modify the initial
{\it disk} mass.
However, it is not sufficient to change the initial disk mass by the
ratio of the observed and the modeled fluxes.
The reason is that a change in the initial mass also alters the rate of
the collisional evolution, whereas we need the ``right'' flux at a fixed
time instant, namely~--- in the reference model~--- at 350~Myr.
Therefore, to find the mass modification factor we apply scaling rules,
as explained in Appendix~A of \citep{Krivov-et-al-2008}.

The  dust, disk, and initial disk masses in the reference model are given
in the first line of Tab.~\ref{tab:masses}.
The dust mass of $7\times 10^{-3}~M_\oplus$ is by about a factor of two
higher than what was derived by \citet{Su-et-al-2006}.
The actual agreement is even better, because our upper cutoff size of
$1000\mum$ is larger than that of \citeauthor{Su-et-al-2005}
The total mass of the reference disk is about $16~M_\oplus$, which is
85~\% of its initial mass 350~Myr ago when the collisional cascade
started to operate.

\begin{table}[t!]
  \caption{Derived dust masses, disk masses, and initial disk masses for all models}
  \label{tab:masses}
  \begin{center}
  \small
  \begin{tabular}{cccc}
    \tableline\tableline
    runs & $M_{\mathrm{dust}}~[10^{-3}M_\oplus]$  & $M_{\mathrm{disk}}~[M_\oplus]$ & $M_{\mathrm{ini}}~[M_\oplus]$ \\ \tableline
    ref. & $6.63$                                 & $16.3$                         & $18.9$\\ \hline\\[-6pt]
    a1   & $5.62$                                 & $4.21$                         & $4.35$\\
    a2   & $3.67$                                 & $3.05$                         & $3.06$\\[4pt]
    b1   & $5.41$                                 & $19.3$                         & $22.7$\\
    b2   & $7.34$                                 & $14.1$                         & $16.2$\\
    b3   & $6.22$                                 & $17.1$                         & $20.7$\\
    b4   & $7.38$                                 & $14.2$                         & $15.7$\\[4pt]
    c1   & $6.96$                                 & $14.8$                         & $17.1$\\
    c2   & $6.59$                                 & $11.9$                         & $13.6$\\
    c3   & $7.08$                                 & $18.5$                         & $21.5$\\
    c4   & $7.28$                                 & $16.5$                         & $19.1$\\[4pt]
    d1   & $10.9$                                 & $41.5$                         & $47.5$\\
    d2   & $4.74$                                 & $10.5$                         & $12.6$\\[4pt]
    e1   & $4.37$                                 & $3.60$                         & $3.81$\\
    e2   & $8.64$                                 & $51.0$                         & $62.7$\\
    e3   & $4.77$                                 & $5.15$                         & $5.70$\\[4pt]
    f1   & $7.09$                                 & $9.31$                         & $10.5$\\
    f2   & $5.92$                                 & $32.2$                         & $39.4$\\[4pt] \tableline
    fit  & $5.09$                                 & $46.7$                         & $55.5$\\ \tableline
  \end{tabular}
  \end{center}
\end{table}

\subsection{Spectral Energy Distribution}

\begin{figure*}
  \centering
  \includegraphics[width=0.83\linewidth]{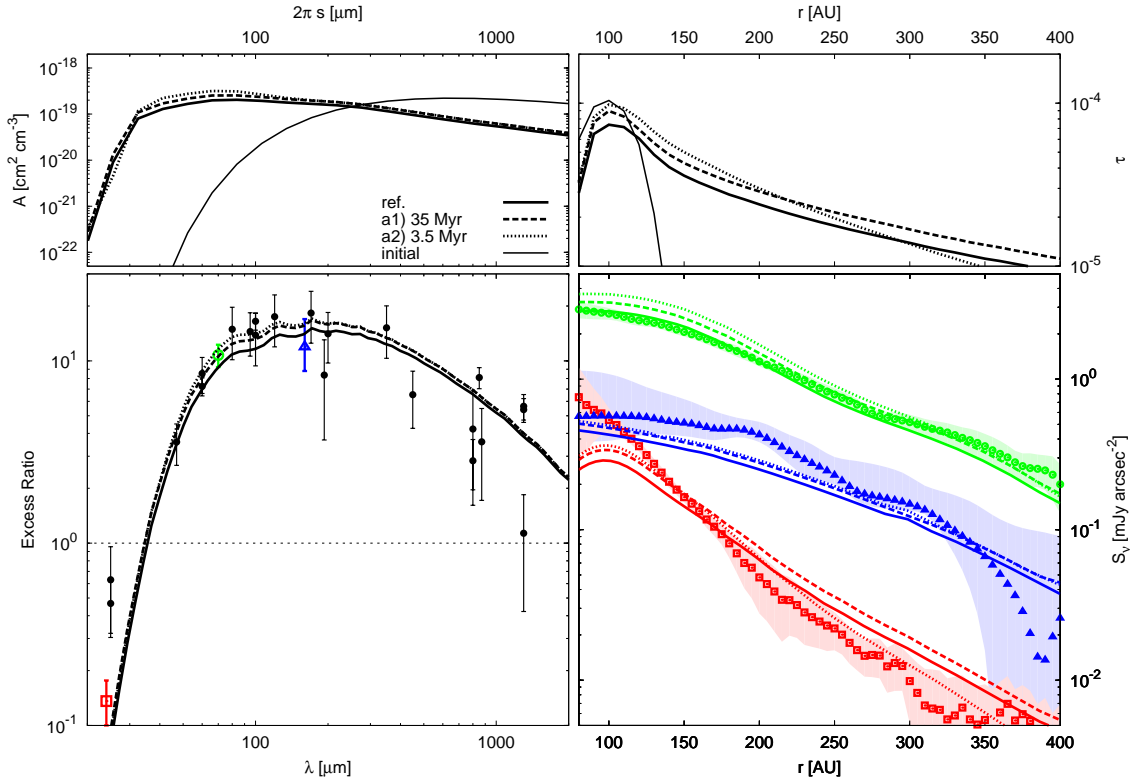}
  \caption{Top left: Grain size dust distributions of the reference model
                     (thick solid line) and of the same model but at
                     earlier times of 3.5 (dotted line) and
                     35~Myr (dashed line), all  in the center of the initial
                     planetesimal ring. The initial distribution for the simulation is 
                     plotted as thin solid line.
                     Note our using $2\pi s$ instead of $s$ in this and
                     subsequent size distribution plots. Since particles with
                     the size parameter $2 \pi s / \lambda \sim 1$  emit most efficiently,
                     $2\pi s$ roughly gives the typical wavelength of the emission.
                     This alleviates comparison between the size distribution and the SED
                     (bottom left).
           Top right: Radial profile of the optical depth for the same
                      disk models.
           Bottom left: Corresponding SEDs.
                        Symbols with error bars are data points
                        (large square, large circle, and large triangle mark 
                        $24\mum$,  $70\mum$, and 
                        $160\mum$ excess ratios deduced from 
                        {\it Spitzer/MIPS}
                        images.
           Bottom right: Modeled (lines) and observed (symbols) surface 
                         brightness profiles at 24, 70, and $160\mum$.
                         The shapes of symbols are the same as 
                         in the bottom left panel and in Fig.~\ref{fig:psf}.
                         The shaded areas
                         around the data points indicate the errors.
          }
  \label{fig:time-mod}
\end{figure*}

For an easier comparison between the modeled SEDs and the photometric
observations in the different spectral regions, throughout the paper we
use the excess ratio.
The latter is defined as the ratio of the dust emission to the stellar
photospheric emission or equivalently, as the ratio of the total flux
(star $+$ dust) to the stellar flux minus unity.
The SED of the reference disk in terms of the excess ratio is presented
in Fig. \ref{fig:time-mod} (left bottom) with a solid line.
Given that our reference model is a first-guess one, the agreement with
the observations is quite satisfactory.

At $24\mum$ the model yields $0.43~\Jy$, which is at $2\sigma$ under the
{\it Spitzer} point of $0.94~\Jy$.
Let us make a more careful comparison, however.
As explained in Sect.~2, the total observed flux from $4''$ outward is
unsure, because of an uncertain part between $4''$ and $10''$.
If we consider only the observed flux from $10''$ outward, it reduces to 
$0.53~\Jy$.
At the same time, we can calculate the flux from $10''$ outward in the
reference model.
The result is $0.40~\Jy$, which is only slightly below the observed one.
That value of $0.40~\Jy$ differs insignificantly from the calculated flux
from $4''$ outward, $0.43~\Jy$, because the emitting dust in the reference
model is entirely located outside $10''$, and it is only the finite PSF's
width that ``transfers'' $0.03~\Jy$ of the emission closer in.

As far as the {\it IRAS} $25\mum$ flux is concerned, it is known to be
quite uncertain due to the large field of view and it is inconsistent
with the {\it Spitzer} $24\mum$ measurement anyway.

In the far-IR the model emission matches the {\it Spitzer} fluxes
perfectly and lies within the error bars of the other observations.
At (sub-)millimeter wavelengths, the measurements themselves are
contradictory, lying sometimes at more than $2\sigma$ from each other,
and it is difficult to judge which of them are most accurate.
The model SED provides a compromise, lying in the middle of the entire
set of data points.

\subsection{Radial Surface Brightness Profiles}

Using {\it SUBITO}, we calculated the radial surface brightness profiles
of our disk model and convolved them with the corresponding PSFs.
The final profiles are presented as solid lines in the right panel of
Fig.~\ref{fig:time-mod}.
The model profiles are not inconsistent  with the {\it Spitzer}
observations.
Especially the $70\mum$ profile is very close to what was measured.
However, both 24 and $160\mum$ curves are slightly too flat.
The $160\mum$ profile lies under the measurements  in the inner part of
the disk and above them in the outer part, explaining  why the total
$160\mum$ flux is about right (see the bottom left panel).
In contrast, most of the $24\mum$ flux comes from the inner part of the
disk inside 100~AU.
In this region the model profile goes below the data points, so that the
higher emission in the outer part of the disk cannot compensate this
deficiency.

Altogether, we state that the surface brightness profiles are more
constraining for the disk model than the SED.
Already our ``first-guess'' model satisfactorily reproduces the observed
SED, but the brightness profiles reveal moderate deviations from those
deduced from the observations.

\section{VARIATION OF MODEL PARAMETERS} \label{sec:improvements}

In this section we investigate how the observables (SED, brightness
profiles) respond to changes in physical parameters (those of the star,
planetesimal belts, and dust, as well as the collisional prescription).
A specific goal is to check if we can improve the agreement of the
modeled brightness profiles of the Vega disk with observations,
preserving the agreement in the SED that we achieved in the reference
model.
Accordingly, we consider a set of models, the parameters of which are
listed in Tab. \ref{tab:parameters}.
Most of these models differ from the reference model by one parameter.
We modify several parameters at a time only if these are physically
related and this is required for consistency.

For each of the models, we present the results in the same way as for the
reference one.
The size distribution, optical depth profile, the SED, and the radial
brightness profiles are combined into a single figure
(Figs.~\ref{fig:loc_i-mod} to \ref{fig:best-fit}), each having the same
structure as Fig.~\ref{fig:time-mod}.
In all the figures, the reference model is overplotted with a solid line.
The resulting final dust masses, final disk masses, and initial disk masses
are given in Tab. \ref{tab:masses}.

In the following subsections the variations of the reference model are
explained and discussed.
They are structured according to the underlying physical and
astrophysical mechanisms at work.

\subsection{Delayed Stirring} \label{sec:time}

Before a debris disk starts to evolve in a steady-state regime, a
collisional cascade has to ignite and operate for sufficient time.
Initiation of the cascade requires a mechanism to stir the disk
\citep{Wyatt-2008}.
This can be self-stirring by largest planetesimals
\citep{kenyon-bromley-2004a} or stirring by planets orbiting in the inner
gap of the disk \citep{Wyatt-2005b,Mustill-Wyatt-2009}.
External events such as stellar flybys can also stir the disk
sufficiently, it may have been the case for Vega $\sim 5$~Myr ago
\citep{Makarov-et-al-2005}.
Furthermore, after the onset of the cascade, the system needs enough time
to reach a steady-state collisional regime at dust sizes
\citep[e.g.][]{Loehne-et-al-2008}.
Thus the duration of a steady-state disk evolution is generally shorter
than the system's age.
We do not know which particular mechanism may have triggered the cascade
in the Vega disk and how long it is already at work.
It may have started either shortly after the primordial gas dispersal or
much later in the Vega history.

To investigate the effect of the unknown ``collisional age'' of the Vega
disk, we simply took our reference disk model and calculated the SEDs and
surface brightness profiles at earlier time steps.
Fig. \ref{fig:time-mod} shows the results at 3.5, 35 and 350~Myr.
At earlier times, the maximum of the size distribution is slightly more
pronounced (top left).
This results in a moderate enhancement of thermal emission between 50 and
$500\mum$, which is still in agreement with the observations (bottom
left).
The optical depth profile (top right) shows that 3.5~Myr of collisional
evolution is not sufficient to bring enough particles on highly eccentric
orbits, so that the profile is steeper than in the reference model.
The $24\mum$ and $70\mum$ profiles steepen (bottom right), the latter
being no longer compatible with observations.
At $35$~Myr, the optical depth profile is only shifted vertically
compared to the reference model, which indicates that the spatial
distribution has already reached an equilibrium (top right).
The final and initial disk masses (Tab.~\ref{tab:masses}) are close to
each other, which is natural as younger disks have spent less material in
collisions.
Besides, the estimated total masses of younger disks are smaller than in
the reference model.
The reason is that younger disks that are not in a steady-state regime
yet are ``dustier'' than older disks of the same total mass
\citep{Krivov-et-al-2006}.

An overall conclusion is that at least several tens of Myr of collisional
evolution seem to be required to make observables consistent with
observations.

\subsection{Disk Location} \label{sec:location}

Our choice of the initial disk extension in the reference model comes
from resolved images in the sub-mm and radio.
However, a low resolution of these observations still leaves room for
reasonable modifications.
Hence we now vary the initial semimajor axis range of planetesimals,
intentionally pushing them to the limits posed by the images, in order to
see the effects more clearly.

\begin{figure*}
  \centering
  \includegraphics[width=0.83\linewidth]{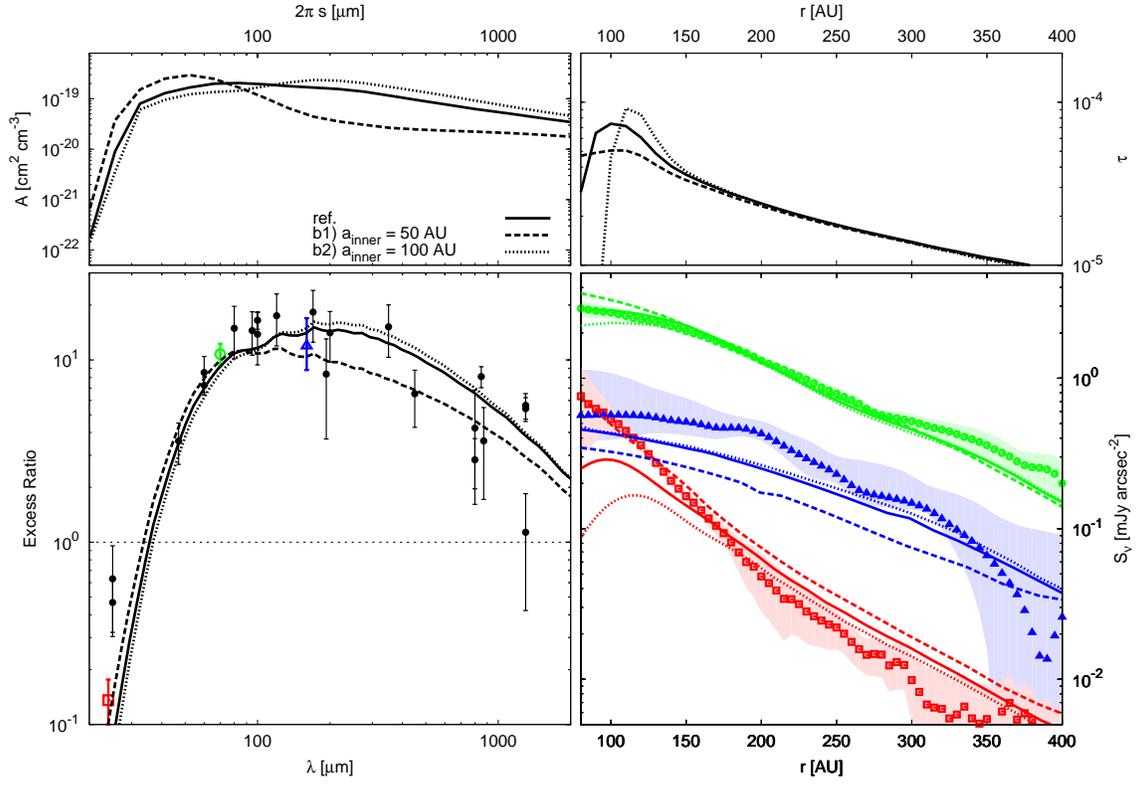}
  \caption{Same as Fig. \ref{fig:time-mod}, but with the inner edge
           of the disk shifted inward (dashed lines) and outward (dotted).
          }
  \label{fig:loc_i-mod}
\end{figure*}

We start with the {\em inner disk edge}.
By placing additional material closer in, one may expect to increase the
warm emission and prevent the brightness profile from dropping off
towards the star too early.
Thus, we try shifting the inner edge in to 50~AU.
For completeness, we also add the case with the inner edge at 100~AU.
The results are presented in  Fig. \ref{fig:loc_i-mod}.

As expected, taking $\ain = 50$~AU slightly shifts the SED to shorter
wavelengths and strongly depresses the sub-mm emission.
Taking $\ain = 100$~AU  yields the opposite, although the effect is
weaker.
This traces back to the under- or overabundance of larger grains,
hundreds of micrometers in size (top left).
The reason for that, in turn, is the existence of two distinct dynamical
regimes for bound dust grains.
Large grains with no or little response to the stellar radiation pressure
essentially inherit their orbital eccentricities from the parent bodies.
For small, barely bound grains, radiation pressure is the dominant
effect, pushing them to wide, highly eccentric orbits.
Shifting the disk further in increases the average collisional
velocities, lowers the collisional lifetime of larger grains -- but not
of smaller ones.
As a result, the relative abundance of larger grains is reduced.

In terms of the surface brightness profiles, it is mostly the $24\mum$
profile that is affected.
It rises significantly inside 200~AU, reproducing perfectly the
observations.
However, in the outer part of the disk it becomes flatter so that the
emission here is overestimated. The $70\mum$ profile remains almost
unaffected, except for the inner part within 150~AU, which responds to
$\ain$ in the same manner the $24\mum$ does, albeit less strongly.
Finally, the $160\mum$ profile preserves its slope, but shifts downwards
($\ain = 50$~AU) or slightly upwards ($\ain = 100$~AU).

\begin{figure*}
  \centering
  \includegraphics[width=0.83\linewidth]{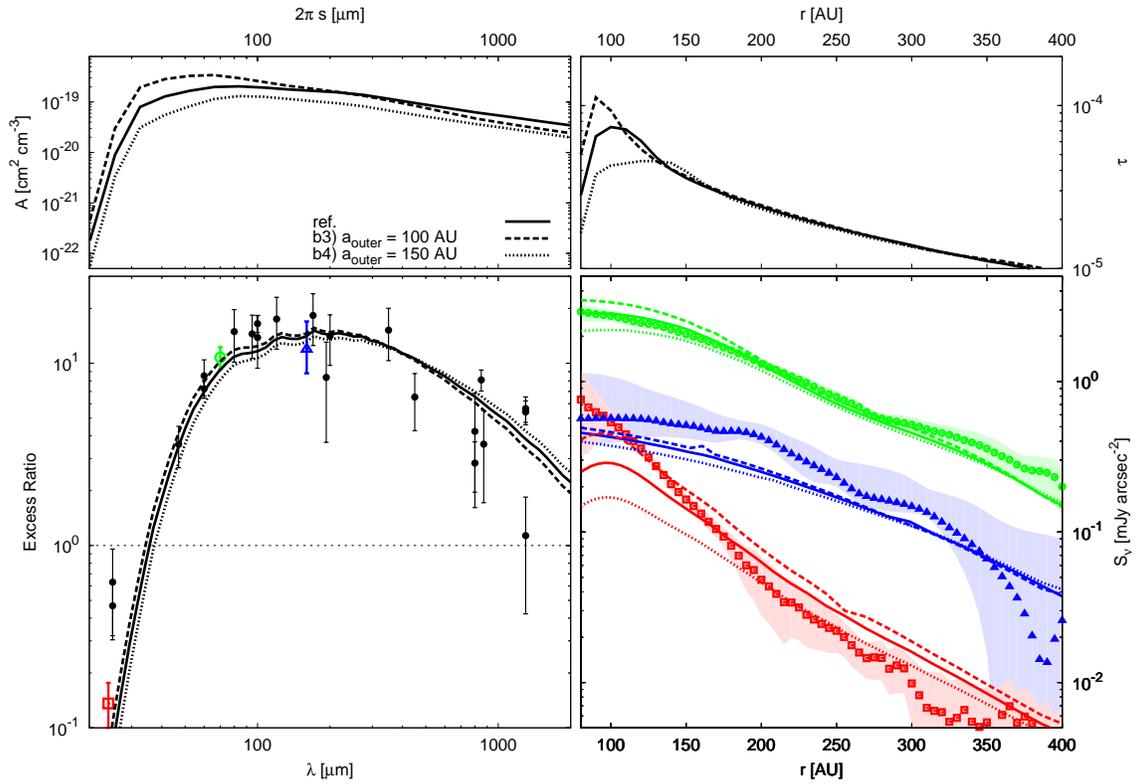}
  \caption{Same as Fig. \ref{fig:time-mod}, but with the outer edge
           of the disk shifted inward (dashed lines) and outward (dotted).
          }
  \label{fig:loc_o-mod}
\end{figure*}

Similar to the inner edge, an inward shift of {\em the outer edge} would
lower the amount of cold dust, enhancing the warm emission.
So we changed $\aout$ to 100 and 150~AU to find similar modifications in
the dust distribution and thermal emission as above
(Fig.~\ref{fig:loc_o-mod}).
Decreasing $\aout$ makes the ring narrower and shifts the bulk of the
material closer in.
The maximum in the size distribution becomes more pronounced and shifts
to smaller grains.
The entire SED slightly shifts towards shorter wavelengths.
The peak of the optical depth profile becomes stronger and moves closer
to the star.
This directly translates to the radial surface brightness profiles,
especially at $24\mum$, which becomes appreciably steeper.
Increasing $\aout$ naturally has opposite effects.

On the whole, it seems that shifting the inner edge of the belt inward
has a clear potential of getting the $24\mum$ profile that would better
match the observed one.
However, the shift from $80$~AU down to $50$~AU that we have tested may
be too strong, because it may contradict to the sub-mm images.

\subsection{Dynamical Excitation} \label{sec:excitation}

We now consider the dynamical excitation of the disk, as parameterized by
the maximum orbital eccentricities $\emax$ that planetesimals had at the
onset of the collisional cascade.
This value does not change considerably in the course of the subsequent
evolution (under the assumptions of our model, e.g. without planets), and
it is approximately the same for disk solids of all sizes except for the
smallest dust particles that are vulnerable to radiation pressure.
From  the dynamical point of view, higher eccentricities increase the
collisional velocities (although the collisional rates remain nearly the
same, see, e.g. \citeauthor{Queck-et-al-2007}\citeyear{Queck-et-al-2007})
and thus the efficiency of the collisional cascade.

We ran two simulations: one with reduced ($\emax = 0.1$) and one with
increased eccentricity ($\emax = 0.3$).
The inclination of the disk was taken to fulfill the energy equipartition
condition, $\imax = \emax/2$.
Since changing $\emax$, but maintaining the same distribution of
semimajor axes would change the radial extension of the disk, we chose to
alter $\ain$ and $\aout$ in such a way as to preserve the radial
extension of the reference model disk.
(Strictly speaking, all this applies to the {\em initial} disk, because
we have no control over the disk extension at later times in the course
of its dynamical evolution.)

\begin{figure*}
  \centering
  \includegraphics[width=0.83\linewidth]{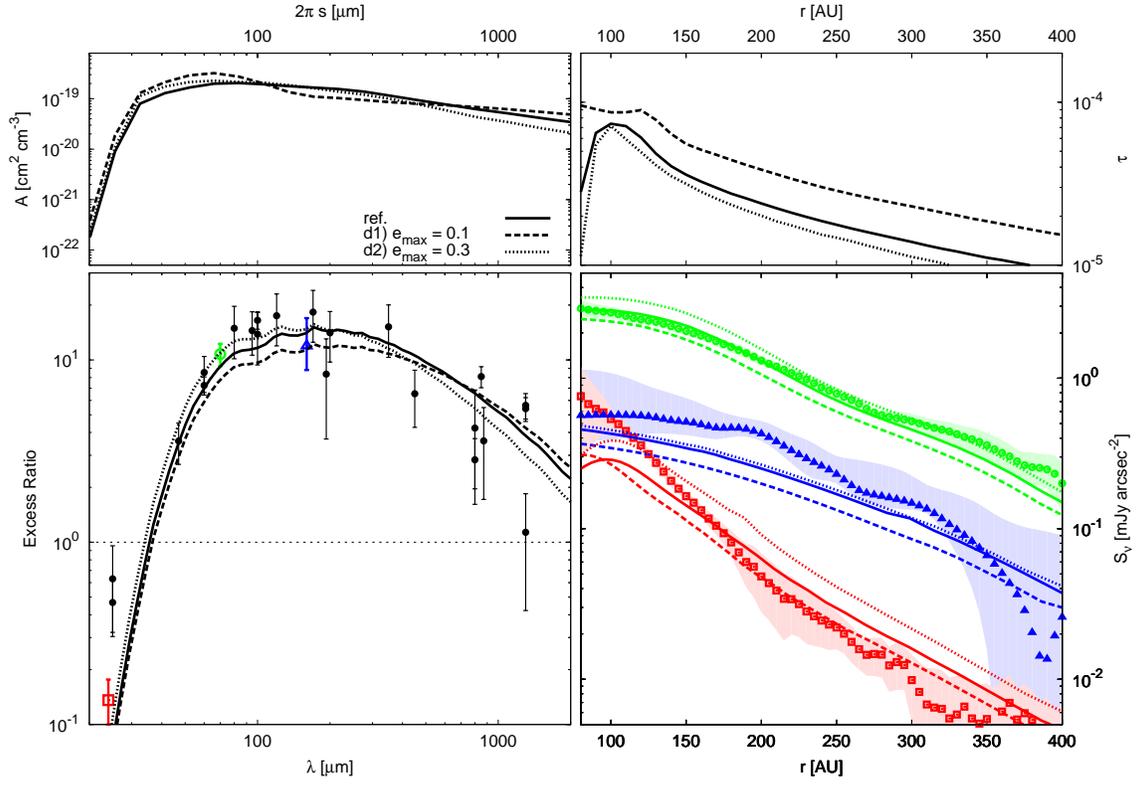}
  \caption{Same as Fig. \ref{fig:time-mod}, but for 
           dynamically less (dashed lines) and more (dotted lines)
           excited model disks.
          }
  \label{fig:ecc-mod}
\end{figure*}

The size distribution in Fig. \ref{fig:ecc-mod} shows that higher
eccentricities lead to a depletion of larger grains ($> 30\mum$) and in
return to an overabundance of smaller grains close to the blowout size.
Consequently, the SED drops beyond $\approx 200\mum$ and rises at shorter
wavelengths becoming more narrow.
In contrast, with lower eccentricities more grains of radii $> 100\mum$
survive, so that fewer particles with $15\mum < s < 100\mum$ are created
in collisions, which results in a more pronounced maximum between the
blowout and $30\mum$.
This yields an enhancement of the radio emission and an overall flatter
shape of the 
SED.

The slope of the optical depth profile in the outer part of the disk
remains nearly unchanged, it just shifts vertically.
However, in the range of the birth ring the optical depth profiles for
different $\emax$ are different:
the lower $\emax$, the broader the maximum.
It is because for higher $\emax$, the distribution of semimajor axes is
narrower and the collisional production of dust near the center of the
ring is much higher than elsewhere.
For lower $\emax$, the semimajor axes are distributed more broadly and
dust is collisionally produced at comparable rates everywhere in the
ring.

The brightness profiles respond to the changes in the disk excitation in
a similar way.
In the outer part they just shift vertically, and most of the changes are
in the birth ring region.
When $\emax$ is reduced to $0.1$, in the outer disk the $24\mum$ profile
matches the observed profile closely.
In the region of the birth ring it is still too low, as is the reference
profile.
However, the emission now keeps rising inward down to $80$~AU, i.e. all
the way through the birth ring, as does the observed emission.

\subsection{Stellar Luminosity} \label{sec:lum}

Changing the stellar luminosity has a two-fold effect on the results.
First, it alters the $\beta$-ratio of the dust grains, affecting their
dynamics.
Second, a different luminosity affects the temperature of the dust
grains, thereby changing the SED and brightness profiles.
Note that all these changes influence only the dust portion of the disk,
not the larger objects.

As mentioned above, Vega is a rapid rotator and so the radiation flux
emitted from its surface varies with the stellar latitude.
In the reference model, we adopted the ``equatorial luminosity'',
$28~L_\odot$.
However, the dust disk ``sees'' not only the stellar equator, but also
receives stellar radiation from higher altitudes.
Thus, we now test the average luminosity of $37~L_\odot$ as derived by
\citet{Aufdenberg-et-al-2006} and, as an extreme case, the canonical
polar value of $57~L_\odot$ (used by many modelers before).

\begin{figure*}
  \centering
  \includegraphics[width=0.83\linewidth]{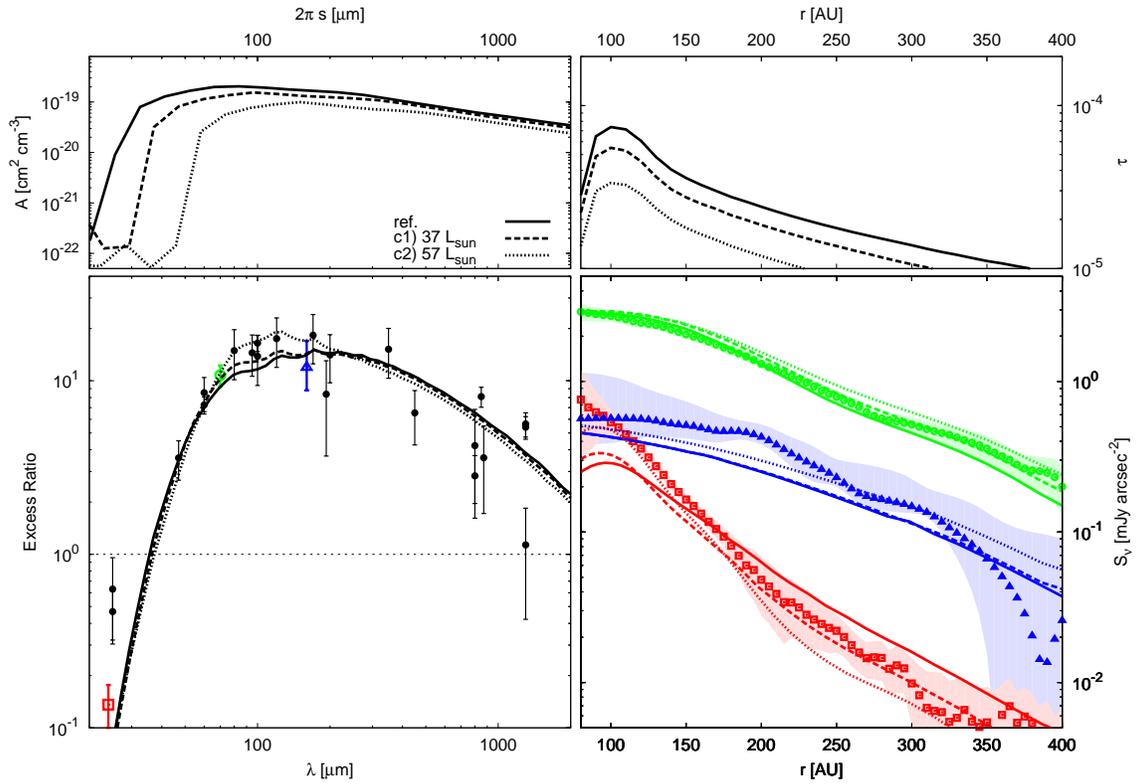}
  \caption{Same as Fig. \ref{fig:time-mod}, but assuming a more luminous central star with
           luminosity of 37 (dashed lines) and 57~$L_\odot$ (dotted lines).
          }
  \label{fig:lum-mod}
\end{figure*}

As the luminosity gets higher, the blowout size increases
(Fig.~\ref{fig:beta}, right), reaching $10\mum$ for a $57~L_\odot$
central star.
The entire size distribution shifts horizontally towards larger sizes and
the jump at the blowout radius becomes more abrupt
(Fig. \ref{fig:lum-mod}).
The optical depth profile preserves its shape, but moves downward.
The reason for the decrease of the optical depth level at higher
luminosities is simply the increase of the grains' blowout size, so
that there are fewer grains that could stay in bound orbits.

Interestingly, the SED in Fig. \ref{fig:lum-mod} (bottom left) does not
change substantially, as one would expect from dramatic changes in the
size distribution.
The reason for moderate changes seen in the figure can be found by
analyzing the size distribution (Fig.~\ref{fig:T+dist}, top right) and
the temperature plot (Fig.~\ref{fig:T+dist}, top left).
Since for a larger $L$ the size distribution is shifted to larger
particles, the temperature range of the smallest particles, which affects
the outer part of the disk the most, becomes narrower.
Consequently, the SED slightly narrows, too,  and the maximum at
$\sim 100\mum$ becomes more pronounced.

The  changes in the surface brightness profiles are two-fold.
While the $24\mum$ profile steepens with increasing luminosity, the 160
and especially the $70\mum$ curves flatten.
The explanation for this behavior is as follows.
The farther out from the star, the faster the temperature decreases with
increasing grain size (Fig.~\ref{fig:T+dist}, top left).
A comparison with the position of the maxima in the size distribution at
different distances and for different luminosities
(Fig.~\ref{fig:T+dist}, top right) demonstrates that an average
temperature in the region of the birth ring increases with increasing 
$L$.
In the outer disk the effect is reverse:
far from the star, a higher stellar luminosity lowers the typical dust
temperatures.
These effects explain why for higher luminosities the mid-IR emission
rises in the inner disk and drops in the outer one, steepening the
$24\mum$ profile.
At the same time, the far-IR emission becomes more efficient farther out,
which flattens the 70 and $160\mum$ profiles.

Our analysis definitely favors an intermediate value of the Vega
luminosity, exemplified by $L = 37~L_\odot$ in our tests.
First, this choice is well justified physically.
Indeed, dust is exposed to stellar light coming from a range of
latitudes, thus the ``right'' luminosity should be between the equatorial
and polar one.
Second, it does provide a better agreement with observations.
Changes in the 70 and $160\mum$ profiles are only marginal, so that they
still match the observations well enough, while the $24\mum$ profile
steepens inside $\approx 250$~AU, coming much closer to the observed
profiles.

\subsection{Chemical Composition} \label{sec:chem}

Like the stellar luminosity, the chemical composition of grains affects
both the $\beta$-ratio (through the radiation pressure efficiency and bulk
density) and dust temperatures (through the absorption efficiency).

Mid- and far-IR spectra of some debris disks reveal distinctive features
\citep{Jura-et-al-2004,Chen-et-al-2006}, which allows one to get insight
into the mineralogy of the dust grains.
For example, spectra of several disks were matched by a mixture of
amorphous and crystalline silicates, silica, and several other species
\citep{Schuetz-et-al-2005,Beichman-et-al-2005b,Lisse-et-al-2007,%
Lisse-et-al-2008a}, including possibly water ice \citep{Chen-et-al-2008}.
Unfortunately, the spectra of the Vega disk (available in the
{\it Spitzer} archive) do not exhibit unambiguous features, which poses
no observational constraints on its composition.

\begin{figure*}
  \centering
  \includegraphics[width=0.83\linewidth]{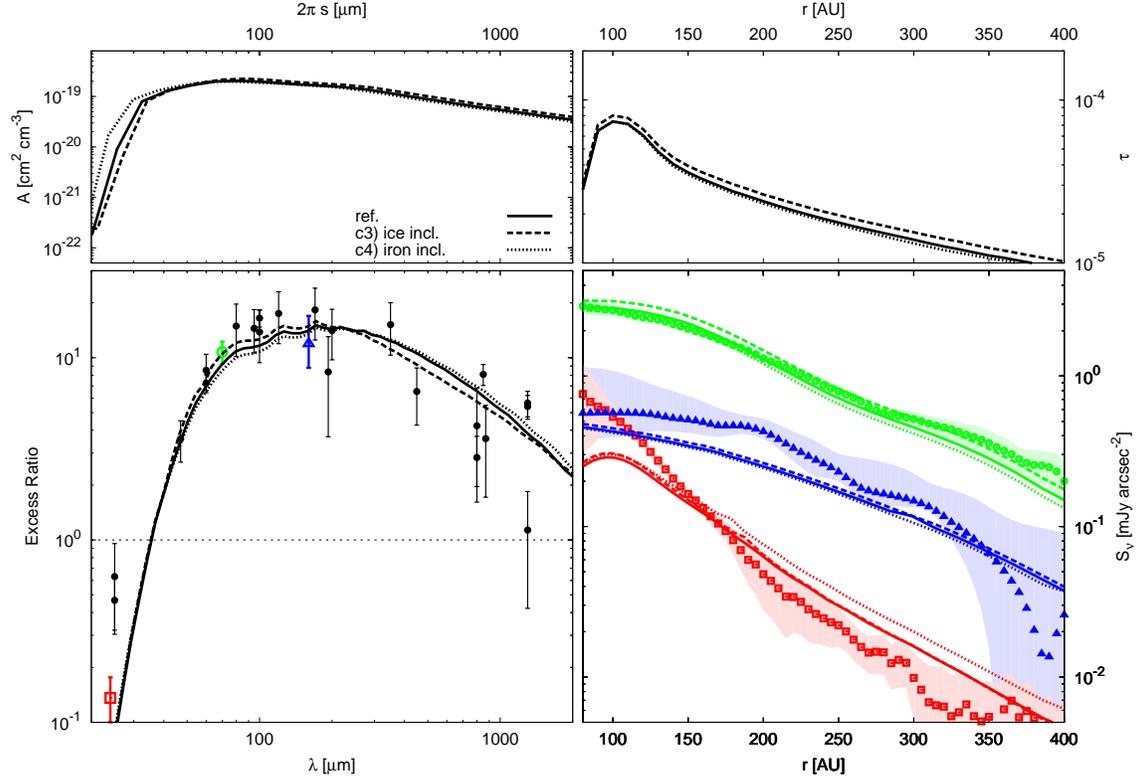}
  \caption{Same as Fig. \ref{fig:time-mod}, but for disk of particles
           consisting of an astrosilicate matrix with 10~\% water ice
           (dashed lines) and iron inclusions (dotted lines).
          }
  \label{fig:chem-mod}
\end{figure*}

In the reference model we used pure astronomical silicate.
Now, to test possible effects of chemical composition, we consider an
astrosilicate matrix with water ice \citep{Warren-1984} and iron
\citep{Lynch-Hunter-1991} inclusions.
The refractive indices were calculated according to the Maxwell-Garnett
theory.
The amount of inclusions was limited to 10~\%, which is an upper limit
for which the effective medium theory still provides accurate results
\citep{Kolokolova-Gustafson-2001}.
The resulting bulk densities for the mixtures with ice and iron are
$3.062~\g/\cm^3$ and $3.757\g/\cm^3$, respectively (compared to
$3.3~\g/\cm^3$ for pure astrosilicate).
The effect of inclusions on the absorption efficiencies (except for small
grains in the near-IR) and $\beta$-ratio (Fig. \ref{fig:beta}) is minor.
Most of the difference in the size distributions
(Fig. \ref{fig:chem-mod}, top left) probably comes from the changes in
the bulk density.
The blowout for grains with iron inclusions is slightly shifted to
smaller sizes und the whole distribution is stretched, while in the case
of water inclusions the opposite is true.
The radial distribution of dust (top right) remains virtually the same.

In terms of thermal emission, the influence of inhomogeneity is very
weak, too.
Consistently with the modifications in the size distribution, the SED is
narrower for ice and slightly broader for iron inclusions.
The surface brightness profiles retain their overall shape.
Only the $24\mum$ emission in the outer disk becomes slightly stronger
and more gently sloping, when iron inclusions are present.

Our conclusion is that inclusions at a 10~\% level have only minor effect
on the observables.
We cannot exclude that more radical changes in the composition would
affect the results substantially, but there is currently no observational
evidence that would justify such changes.

\subsection{Cratering Collisions} \label{sec:cratering}

We turn to an analysis of the underlying collisional model implemented in
{\it ACE}.
The detailed physics and outcomes of binary collisions under the
conditions of debris disks are poorly known, which represents one of the
major sources of uncertainty in our simulation results.
Accordingly, in this and subsequent sections, we vary three key
parameters that control the treatment of collisions.

We first explore a hypothetical collisional cascade, in which only
disruptive collisions operate and the cratering collisions do not occur.
This means that we only consider collisions with specific impact energies
above the threshold value ($Q\sbs{D}^\star$), which shatter both
colliders completely.
All collisions at lower energies (that would in reality erode one or both
of the colliders) are simply ignored.

\begin{figure*}
  \centering
  \includegraphics[width=0.83\linewidth]{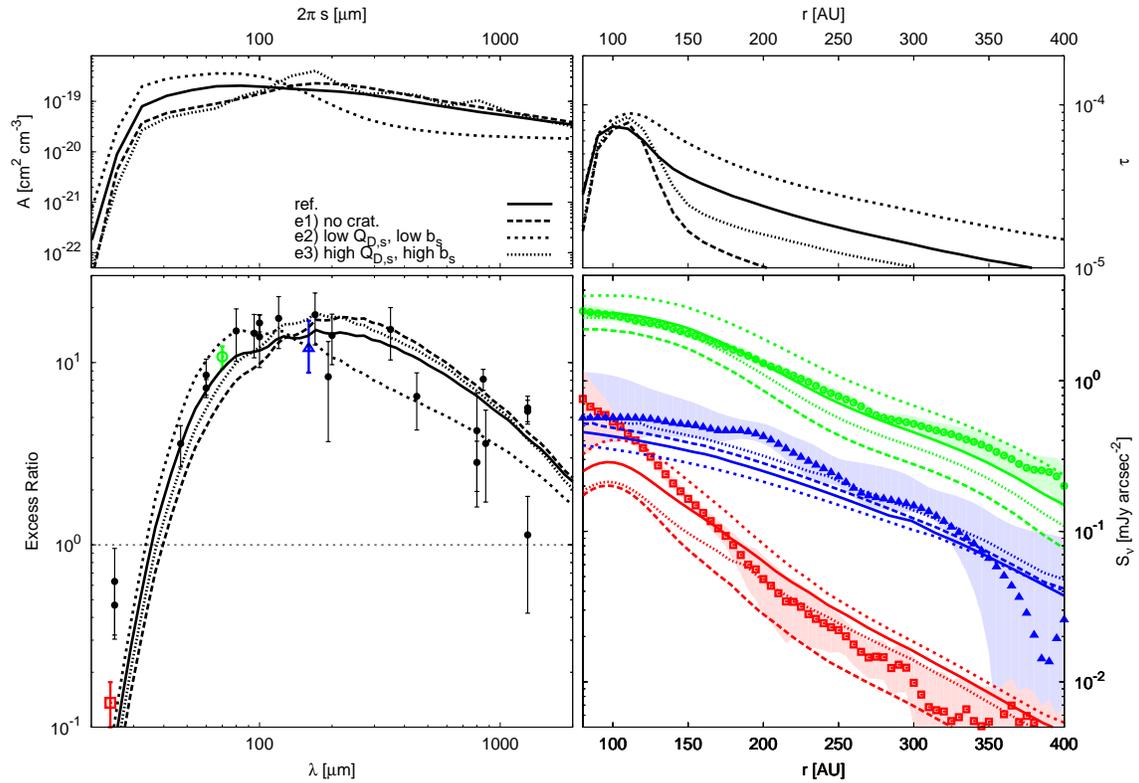}
  \caption{Same as Fig. \ref{fig:time-mod}, but for a disk without
           cratering collisions (long dashed) and for a disk material with lower (short
           dashed) and higher critical shattering energy (dotted).
          }
  \label{fig:crat-mod}
\end{figure*}

Thus, an efficient way of eroding larger objects by collisions with much
smaller grains is switched off.
As a result, grains with $20\mum < s < 300\mum$ are more abundant than
in the reference model (Fig.~\ref{fig:crat-mod}).
And conversely, the number of grains with $s < 20\mum$ decreases, so that
the maximum of the size distribution is now effectively shifted to about
$30\mum$.
The explanation is simple.
Excluding cratering collisions prolongs the collisional lifetime of
larger grains, because smaller impactors that cannot disrupt but would
efficiently erode them, now leave them intact
\citep[see, e.g.,][and references therein]{Thebault-Augereau-2007}.

The change in the size distribution shifts the SED towards longer
wavelengths and slightly narrows it.
There is now  a lack of emission in the mid and far-IR up to $100\mum$
and an excess emission at sub-mm.
The resulting SED clearly violates the data.

Given the deficiency of small particles on highly eccentric orbits, it is
not surprising that the optical depth profile becomes very steep.
As a consequence, the $70\mum$ brightness profile steepens significantly.
In addition, all three profiles fall too low. Like the SED, they are no
longer consistent with the observations.

We conclude that cratering collisions cannot be ignored.
They seem mandatory to reproduce the observations of the Vega disk with
collisional simulations.

\subsection{Energy Threshold for Fragmentation} \label{sec:threshold}

In this subsection, we explore the role of the (unknown) tensile strength
of the solids, parameterized by the shattering energy 
$Q\sbs{D}^\star$ in the strength regime.
To this end, we decrease $Q_{D,s}$ and $b_s$ in Eq.~(\ref{eq:QD}) in one
simulation (``weak material'') and increase them in another one (``hard
material'').

Figure \ref{fig:crat-mod} shows that larger grains benefit from an
increase of the energy threshold in a similar way they do from neglecting
cratering collisions or from lowering the average impact energies.
Their collisional lifetime becomes longer and their amount increases.
And conversely, a lower $Q\sbs{D}^\star$ reduces the amount of larger
particles.
For smaller grains, the number of potentially hazardous impactors is not
determined by the change in the critical impactor mass for disruption
(that comes along with the change in critical energy) but by the blowout
limit, which remains unchanged.
Consequently, decreasing $Q\sbs{D}^\star$ ``supports'' smaller grains,
producing a more pronounced first maximum in the size distribution.

In the SED, a lower critical energy leads to a strong shift of the
maximum to about $80\mum$, and makes the rise in the mid- to far-IR
steeper, whereas the sub-mm and millimeter part lowers and flattens.
The opposite changes, albeit less pronounced, are seen for a higher
critical energy.
In both cases, the agreement between the modeled and observed SED becomes
rather worse.

Similarly to the size distribution, the optical depth profile responds to
a harder material in nearly the same way as to excluding the cratering
collisions.
As far as the surface brightness profiles are concerned, the only real
improvement can be found in the $160\mum$ profile for larger $Q_{D,s}$
and $b_s$.
However, this is accompanied with a steepening and flattening of the 70
and $24\mum$ profiles, respectively, which are then clearly inconsistent
with the observations.

We conclude that using ``weak'' or ``hard'' material with respect to the
nominal one does not generally improve agreement with the observations.
When improving one of the three surface brightness  profiles, for
instance, this makes one or two of the others worse.
We find, however, that results are very sensitive to the critical energy.
Thus moderate modifications in the critical energy can be useful for
``fine-tuning'' of the models.

\subsection{Fragment Distribution} \label{sec:frag_distr}

One more essential part of the collisional description is the
distribution of fragments produced in a single collision.
In the reference model we assumed their mass distribution to follow a
power-law with an index $\eta = 1.833$ retrieved from experiments
\citep{Fujiwara-et-al-1977}, but the experimental conditions do not
necessarily reproduce the conditions of debris disks.
Here, we try another two disk models, one with an enhanced production of
small particles ($\eta = 1.95$) and one with a reduced production
($\eta = 1.6$).

\begin{figure*}
  \centering
  \includegraphics[width=0.83\linewidth]{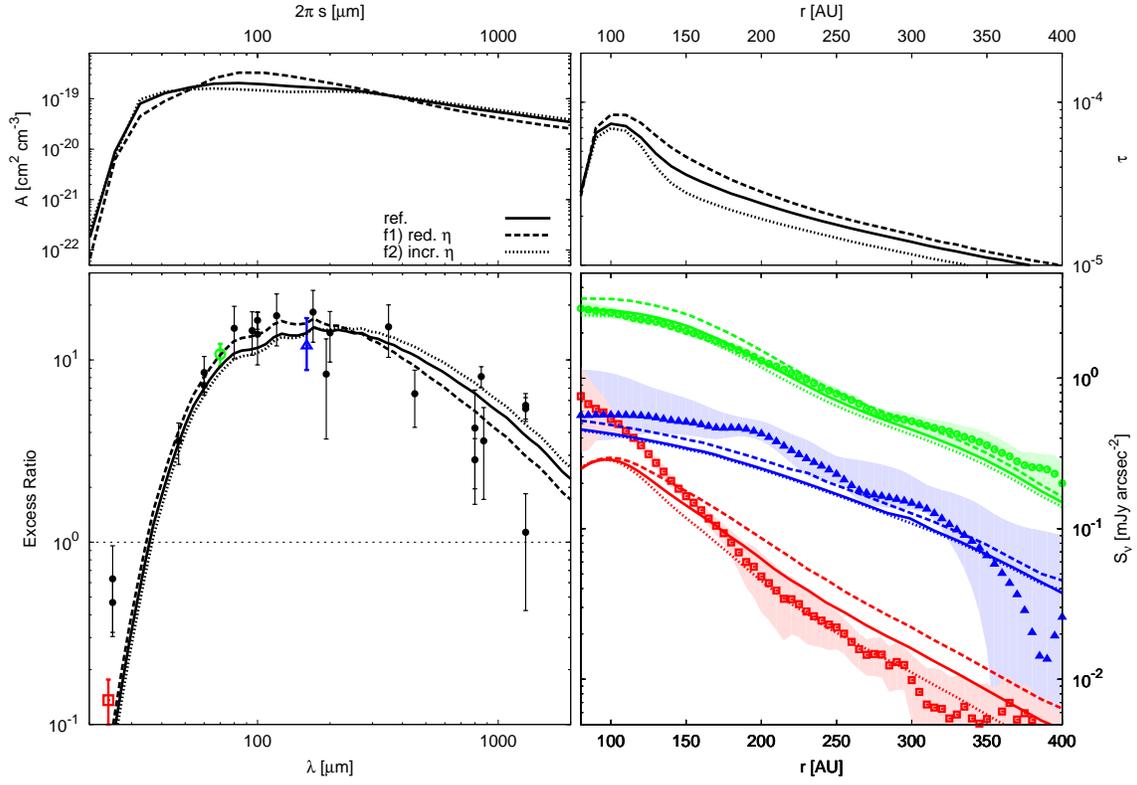}
  \caption{Same as Fig. \ref{fig:time-mod}, but for model disks with a flatter
           (dashed) and steeper distribution of fragments (dotted).
          }
  \label{fig:col-mod}
\end{figure*}

The effect on the size distribution in Fig. \ref{fig:col-mod} is not very
strong.
An increase of $\eta$ enhances the production of small particles so that 
the total distribution becomes flatter.
This makes the SED broader:
the far-IR emission decreases while the sub-mm fluxes are enhanced.
Reducing $\eta$, however, trims the production rate of small particles,
so that the maximum in the size distribution becomes broader and is
shifted to about $15\mum$.
Consequently, the SED becomes somewhat narrower, with a steeper rise in
the mid-IR, stronger emission in the far-IR, and a steeper fall-off in
the sub-mm.
These changes are minor, so that the SEDs for both $\eta$ values are
consistent with the observed SED, as is the SED in the reference model.

Changes in the optical depth are subtle, but not unimportant,
particularly at the outer edge of the birth ring.
While a lower $\eta$ makes the optical depth profile smoother, a higher
$\eta$ creates a slight dip at the outer end of the planetesimal belt.
In this radial zone, emission stems predominantly from intermediate-sized
grains, which are placed by radiation pressure in moderately eccentric
orbits, but still cannot reach the outermost regions of the disk.
Reducing $\eta$ increases the amount of these particles compared to the
reference model, which smoothens the optical depth profile in this
region.
And conversely, an increase of $\eta$ depresses the population
intermediate-sized grains, causing the dip.

In the radial surface brightness profiles these changes are evident in
the $24\mum$ profile, which becomes flatter for small $\eta$ and steeper
for high $\eta$ in the region up to about 200~AU.
The vertical shifts are in agreement with the modifications in the SED.

A conclusion is that a flatter size/mass distribution of fragments with
$\eta = 1.95$ makes the $24\mum$ brightness profile more consistent with
the observed ones, without making other two profiles and the SED worse.

\subsection{The Best Fit \label{sec:best_fit}}

Different modifications in the previous section have shown no simple way
of further improving the agreement of our reference model with the
observations.
However, we found that variation of some parameters is able to change the
results in the desired direction.
We now combine several of the modifications that looked promising:
the disk was extended inwards, the eccentricities were reduced, the
luminosity increased, and a steeper fragment distribution was assumed.
Specific parameter values are listed in the last line of
Tab.~\ref{tab:parameters}.

\begin{figure*}
  \centering
  \includegraphics[width=0.83\linewidth]{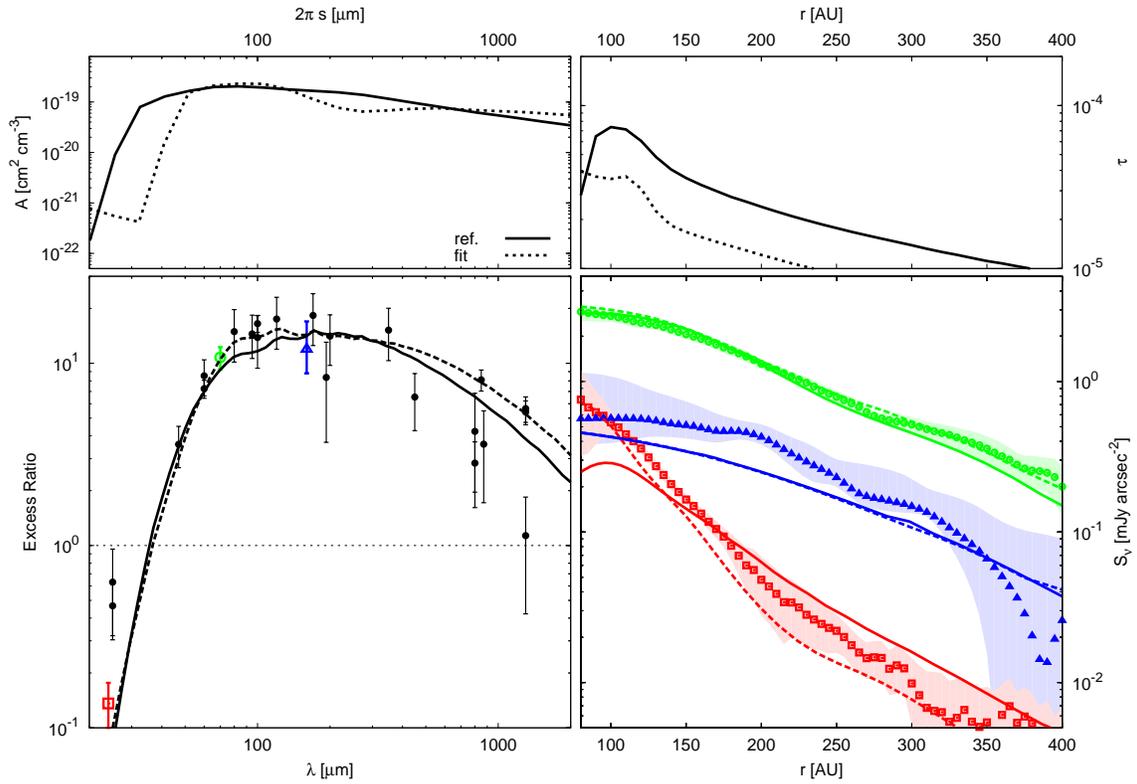}
  \caption{Same as Fig. \ref{fig:time-mod}, but for our best-fit model (dashed lines).
          }
  \label{fig:best-fit}
\end{figure*}

The result is depicted in Fig. \ref{fig:best-fit} with dashed lines;
as always, solid line shows the reference model for comparison.
In terms of the SED (bottom left), all photometry data (aside from the
{\it IRAS} $25\mum$ points) are reproduced within the error bars
shortward of $200\mum$.
At longer wavelengths, the observational data themselves split into
groups that are not in agreement with each other.
The model perfectly matches the upper set of points.

In terms of the radial surface brightness profiles, the $160\mum$ profile
is nearly the same as in the reference model.
The $24\mum$ profile is at about $1\sigma$ in all regions of the disk.
From all three curves the $70\mum$ profile is the closest to the observed
one.

\section{DISCUSSION} \label{sec:discussion}

\subsection{Modeling the Vega Disk ``from the Sources''}

In this paper we have attempted to reproduce observations of the Vega
disk by assuming that the observed emission stems from the debris dust,
which is produced by a ``Kuiper belt'' at $\sim 100$~AU in a steady-state
collisional cascade.
To this end, we have performed involved collisional simulations to
generate and evolve the disk of solid material from planetesimals to dust
and then calculated thermal emission of the dust portion of the disk to
confront the results with the available observations.
Compared to the commonly used modeling technique, in which {\em dust}
distributions that provide the best fit to the data are sought, we seek
{\em parent planetesimals} whose collisional debris could produce
emission that matches the observations.
This is obviously one major step more than in the traditional method:
we try to go back to sources by modeling actual physical processes that
operate in a debris disk.

Starting this research, we wanted to test whether our approach can
withstand a detailed comparison with various sets of observational data
on one particular, and well studied, debris disk system.
We have chosen Vega, an archetype debris disk.
One particular motivation for this choice was that previous studies
uncovered peculiarities, casting doubt on a ``standard collisional
cascade'' as the main mechanism sustaining the disk of Vega.

We understood from the very beginning that our efforts may not be
rewarding.
The fact that our approach involves a physical modeling to the whole
``depth'' automatically adds major uncertainties, notably those related
to the collisional physics.
Contrary to the standard data fitting, we cannot control size and radial
distribution of visible dust and we are not at liberty to add, for
example, another population of dust to improve agreement with
observations.
Thus it would be unrealistic to expect that our approach could
immediately deliver results that are consistent with observations to that
same extent as in the traditional method.
It was to our surprise therefore that already with a `first guess'
reference model we are able to reproduce the SED and the modeled mid- to
far-IR radial brightness profiles were not inconsistent with
{\it Spitzer/MIPS} data.
We were then able to further improve the agreement by variation of model 
parameters.

\subsection{Blowout or bound grains?}

It is interesting to trace, why \citet{Su-et-al-2005} needed an extremely
high amount of blowout grains to explain the {\it MIPS} observations,
whereas we are able to reproduce the same data without any blowout grains
(in our models, they make a negligible contribution to the SED and radial
brightness profiles at all wavelengths considered and at any distance in
the disk).
By fitting the {\it MIPS} photometry and radial profiles with a single
power-law size distribution, \citet{Su-et-al-2005} found the best fit to
be $s_{min}= 1\mum$ and $s_{max} = 50\mum$, with a slope of $-3.0$.
Assuming nominal stellar luminosity of $L = 60~L_\odot$, already the
compact grains have a blowout limit of $\approx 8\mum$.
However, if the grains are highly porous, which of course they may, then
the blowout limit will shift to much larger sizes, so that all grains in
the $1\mum$...$50\mum$ could indeed be in unbound orbits.
An additional argument to favor the radiation pressure-induced outflow
seemed to be the deduced brightness profile at $24\mum$ with a slope of
$-3$...$-4$, since a $-3$ slope is what blowout particles with nearly
constant ``terminal'' velocities would produce.
At that time, it was not yet realized that slopes in the range
$-3$...$-4$ would equally be typical of an extended disk of small bound
grains in elliptic orbits around a parent planetesimal ring, as was found
later numerically 
\citep{Krivov-et-al-2006,Thebault-Augereau-2007} and analytically
\citep{Strubbe-Chiang-2006}.

In this study, we assume compact grains and $L = 28~L_\odot$ in the
reference model (note that the very idea to reduce the stellar luminosity
would sound strange at the time of the \citep{Su-et-al-2005} study,
because the fast rotation of Vega was not yet discovered).
As a result, the blowout radius reduces to $\approx 4\mum$.
Thus where \citeauthor{Su-et-al-2005} had $1\mum$...$50\mum$ grains, we
have $4\mum$...$50\mum$ ones.
For a $-3.0$ size distribution slope, the emitting cross section area is
equally distributed over the sizes.
Thus, even if we take into account that smaller grains are somewhat
hotter than larger ones, by excluding the $1\mum$...$4\mum$ subrange we
do not lose much of the $24\mum$ emission compared to
\citeauthor{Su-et-al-2005}
The radial distribution of dust in our model, as explained above, is not
very different either.
That is why we arrive at a similar level of $24\mum$ emission.
In terms of dust mass, the difference is even smaller.
Indeed, the dust masses we derive here ($6.6 \times 10^{-3} M_\oplus$ in
the reference model, see 
Tab.~\ref{tab:masses}) are close to those derived by
\citeauthor{Su-et-al-2005} ($2.8 \times 10^{-3} M_\oplus$).

\subsection{Mass loss from the disk}

We now make estimates of the mass loss from the disk.
The total mass of dust (up to $1\mm$) in the reference model is
$\sim 7 \times 10^{-3} M_\oplus$ (Table~\ref{tab:parameters}).
Assuming for simplicity that the size distribution follows a $-3.5$ power
law, the mass of the smallest bound grains (say, up to $10\mum$) is
$\sim 7 \times 10^{-3} M_\oplus \times \sqrt{10\mum/1\mm}
\sim 7 \times 10^{-4} M_\oplus$.
The steady-state mass of blowout grains is then by a factor of 100
smaller (a strength of the dip in the size distribution, which is the
ratio of the collisional lifetime of bound grains to the disk-crossing
time of unbound ones, see Fig.~\ref{fig:T+dist} right), giving
$\sim 7 \times 10^{-6} M_\oplus$.
Their lifetime is $\sim 1000\yr$, so that the mass loss rate is
$\sim 7 \times 10^{-9} M_\oplus\yr^{-1}$.
Therefore, over the system's age, 350~Myr, the disk must have lost
$\sim 2 M_\oplus$ of material.
This estimate is consistent with the difference between the initial and
final disk masses given in Tab.~\ref{tab:masses}, typically a few
$M_\oplus$.

\subsection{The $24\mum$ Emission}

It was ``the $24\mum$ problem''~-- an apparently too strong and radially
extended $24\mum$ emission compared to what was expected from dust in
bound orbits~--- that triggered debate on whether the Vega disk contains
an excessive number of small blowout grains, incompatible with a
steady-state collisional cascade \citep{Su-et-al-2005}.
Thus we now discuss in more detail how the $24\mum$ flux predicted by our
models compares to {\it Spitzer} data.

Although the models presented here are in a reasonable agreement with
observations, most of them somewhat underestimate the observed $24\mum$
emission in the parent ring region, at $80$--$120~\AU$ or $10''$--$15''$,
while slightly overestimating it farther out from the star.
As a result, the total flux outside $10''$, which is dominated by the
flux from the ring region, is slightly below the observed value.
For example, our reference model predicts $0.40~\Jy$ outside $10''$, while
the observed flux is $0.53~\Jy$.
These deviations are subtle and probably not of serious concern.
We argue that they may simply be caused by the roughness of the model
(see Sect.~\ref{sec:caveats}). Indeed, we were able to find a combination
of model parameters (Sect.~\ref{sec:best_fit}) which reproduced the
observed $24\mum$ profile outside $10''$ quite well.

However, the question of $24\mum$ emission from the inner system
($<10''$) remains open.
Our analysis in Sect.~\ref{sec:datareduction} yields a total $24\mum$
flux from $4''$ ($30~\AU$) outward of $0.94~\Jy$, although the true value
may be lower, because central part of the {\it MIPS} images is saturated.
Assuming, however, the data to be accurate, the rise of the $24\mum$ flux
from $10''$ inward can hardly be explained with the models presented
here.
A natural explanation would be an additional dusty belt in the system at
$\sim 10~\AU$.
Such a belt could enhance the $24\mum$ emission coming from the ``main''
disk.
As yet it is not clear, however, if any constraints on such a belt can be
found in the {\it Spitzer/IRS} spectrum of Vega.
Nor is it clear whether the inner system may accommodate such a belt if,
as conjectured, it hosts one or more close-in planets.
In the future, this simple hypothesis could be checked or falsified
directly, for instance with mid-IR interferometry.

\subsection{The $850\mum$ Emission}

So far, we confined our analysis to resolved images in the IR and did not
consider explicitly sub-mm and radio images.
The main reason for that is a low resolution of such measurements,
implying that only weak constraints can be put on the radial brightness
profiles at long wavelengths.
However, at least the total flux at sub-mm wavelengths derived from the
images serves as an additional test to the models.

In an analysis by \citet{Su-et-al-2005}, the observed sub-mm emission
could not be reproduced with a two-component dust disk ($2\mum$ and
$18\mum$ in radius) that was sufficient to fit all available data at
shorter wavelengths.
To cope with the problem, they artificially added a population of larger
grains, with a radius of $215\mum$.
In our approach, solids from dust to planetesimals have a continuous size
distribution, which is not postulated, but physically modeled.
From Figs.~\ref{fig:time-mod}--\ref{fig:best-fit} (left bottom panels) it
is apparent that our simulations naturally reproduce the sub-mm flux with
a reasonable accuracy.
An additional consistency check that we made was to calculate the
$850\mum$ profile of our best fit model (Sect.~\ref{sec:best_fit} and 
convolve it with a Gaussian of $16^{\prime\prime}$ beam size.
The resulting profile was then compared with the {\it SCUBA} profile
extracted by \citet{Su-et-al-2005} from the original images published by
\citep{Holland-et-al-1998}.
The modeled profile is slightly narrower than that observed and the
maximum between 50 and 100~AU is by a factor of two lower.
Given the large width of the PSF and calibration uncertainties of
{\em SCUBA} observations, and that Mie calculations likely underestimate
sub-mm emission (as discussed in Sect.~\ref{sec:caveats}), we deem the
agreement with the data satisfactory.

\subsection{Model Parameters}

As we saw, even our ``best fit'' (Fig.~\ref{fig:best-fit}) cannot match
the observations perfectly.
The discrepancies could stem either from the fact that our
``first-guess'' choice of parameters in the reference model was not the
best, from the limitations or shortcomings of the collisional and thermal
emission model, or for the two reasons together.
In this and the next section, we address the both possible reasons in
turn.

In Sect.~\ref{sec:improvements}, we investigated in detail how the dust
distributions and thermal emission are affected by a large array of
physical parameters:
\begin{itemize}
  \item The ``collisional'' age of the disk, i.e. the time elapsed
        from the onset of the steady-state collisional evolution 
  \item The location and extension of the parent planetesimal belt
  \item The dynamical excitation of the belt, parameterized by the maximum
        orbital eccentricity of planetesimals
  \item The stellar luminosity
  \item The chemical composition of the visible dust
  \item Several parameters that control individual collisions between
        the objects in the disk, from planetesimal to dust.
        In particular, we checked the role of cratering collisions, 
        the critical impact energy threshold for disruptive collisions, and
        the distribution of fragments after a collision.
\end{itemize}

We have identified two parameters that have a major influence.
One is the stellar luminosity, and its effect is particularly
interesting.
One might think that a more luminous central star would increase dust
temperatures, but this is untrue.
A higher luminosity implies a larger blowout size.
Thus the most abundant grains in the disk~--- those just above the
blowout limit~--- are now larger and therefore colder
(Fig.~\ref{fig:T+dist}, top left).
The net result is that, counterintuitively, the characteristic dust
temperature in the disk does not change much (in the outer disk it even
decreases) with increasing luminosity of the central star.
Another parameter is the efficiency of cratering collisions.
In the extreme case where these are switched off, the simulation results
contradict the observations.

Contrary to our expectations, it turned out that other parameters probed
have only minor to moderate effect on the SED and surface brightness
distribution of the disk, with consequences being two-fold.
On the one hand, this implies that the model predictions are rather
robust.
As an example, our reference model was already in a rough agreement with
observations, so that further improvements through ``parameter tuning''
seemed easy, but these turned out to be difficult.
We find meaningful parameter sets that provide better fit to observations
than the reference model does (see Fig.~\ref{fig:best-fit}), but the
agreement is somewhat worse than in the study of \citet{Su-et-al-2005}.
On the other hand, a weak dependence of the observables on many
parameters  restricts the possibility of constraining them, which is
somewhat unlucky, because  placing constraints on model parameters is one
of the important goals of the modeling.

\subsection{Limitations of the Model \label{sec:caveats}}

We are fully aware that our modeling approach, as every other, involves
a number of simplifying  assumptions that may limit its applicability and
influence the results. Here, we list the most important caveats.

Many assumptions have been made in describing collisional physics.
Our collisional prescription approximates the critical shattering energy
with two power laws (Eq.~\ref{eq:QD}), which may be particularly crude at
dust sizes \citep[e.g.][]{Thebault-Augereau-2007}.
The mass of the largest fragment and the distribution of smaller debris
may deviate from what was assumed here, and any real disk should be
composed of objects whose mechanical properties (and even the bulk
density) vary from one object to another (e.g. pre-shattered objects
could be less dense and more loosely bound than pristine ones).

A major simplifying assumption in treating the dynamics of planetesimals
and their dust is that we ignore alleged planetary perturbers interior to
the main belt.
The consequences are discussed in Sect.~\ref{sec:planets}.
At dust sizes, we did not take into account P-R drag, its role is
discussed in Sect.~\ref{sec:PR}.
In calculating the radiation pressure force acting on dust grains, we
assumed them to be compact and spherical, thus ignoring possible
non-radial effects \citep[e.g.][]{Kimura-et-al-2002}.
Like mechanical properties, optical properties of dust may vary from one
grain to another, resulting in different response to radiation pressure,
different temperatures, and different thermal fluxes even for like-sized
particles \citep{Krivov-et-al-2006}.
Furthermore, even spherical particles are treated in an approximate way.
We applied Mie theory to model the emission properties.
Although this method is classical and commonly used, it should be treated
with caution.
One particular concern is that Mie calculations probably underestimate
the emission in the sub-mm and radio due to neglected size effects
\citep{Stognienko-et-al-1995,Krivov-et-al-2008}.

\subsection{The Role of the Poynting-Robertson Effect \label{sec:PR}}

Poynting-Roberston (P-R) drag mostly affects smallest particles and thus
emission at shortest wavelengths considered.
The P-R force moves such grains inward,
placing some of them interior to the inner edge of the birth ring.
The warm emission of these particles especially around the inner edge of
the birth ring should increase.

However, with the Vega disk's relatively high optical depth
($8.3\times 10^{-4}$ at $100~\AU$ in the reference model), the
collisional timescales of dust grains are shorter than timescales over
which Poyining-Roberston (P-R) drag causes their appreciable radial
displacement (Fig.~\ref{fig:timescales}).
Thus the Vega disk can be referred to as a collision-dominated, rather
then transport-dominated disk \citep{Krivov-et-al-2000, Wyatt-2005}.
Still, it is useful to check to what extent PR drag may affect the
results in terms of SED and brightness profiles.

To this end, we have switched on P-R drag in our reference model.
Unfortunately, when a drag force, the P-R force in our case, is added to
a collisional model, the mass-time scaling law (Sect.~\ref{sec:ace}) is
no longer valid.
One has to take the initial disk mass that yields a disk with the
``correct'' dust mass after 350~Myr, i.e. the dust mass that gives the
maximum of the SED at the level actually observed.
This means trial and error, i.e. several {\it ACE} runs with different
initial disk masses followed by {\it SEDUCE} and {\it SUBITO} runs.
What makes the modeling even more demanding, is that the presence of a
drag force implies diffusion in the phase space of pericenters and
eccentricities, which slows down each {\it ACE} run appreciably.
We had to perform four {\it ACE} runs, each of which took about 20
core-days CPU time.

The ``right'' dust mass after 350~Myr of evolution with P-R was achieved
when the initial disk mass was set to $20.5~M_\oplus$ (instead of
$18.9~M_\oplus$ in the reference model without P-R), and the final disk
mass was $18.0~M_\oplus$ (instead of $16.3~M_\oplus$ without P-R).
As expected, the influence of P-R on the $160\mum$ and $70\mum$ turned
out to be completely negligible.
At $24\mum$, the emission in the outer disk increases by $\sim 10$~\% and
in the birth ring (at $80~\AU$) by $\sim 60$~\%.
Thus the whole $24\mum$ profile gets somewhat steeper, and agrees with
observations slightly better than the original profile in the reference
model (solid lines in Figs. \ref{fig:time-mod}--\ref{fig:best-fit}).
However, the improvement is only minor, and we conclude that the P-R
effect can safely be neglected in modeling the Vega system.

\subsection{Presumed Planets in the Vega System \label{sec:planets}}

One major caveat not discussed in Sect.~\ref{sec:caveats} is that
our collisional model, implemented in the {\em ACE} code, ignores effects
of a possible planet (or planets) interior to the planetesimal belt.
Below we briefly outline the facts that point to the presence of such
planets in the Vega system and discuss how these perturbers may affect
the observed properties of the debris disk.

Asymmetries in the Vega disk were first discovered by
\citet{Holland-et-al-1998} in a {\it SCUBA} $850\mum$ image, and
subsequent sub-mm and radio observations have confirmed  a clumpy ring
structure.
\citet{Wilner-et-al-2002} introduced the idea of a Jupiter mass planet
trapping dust in mean-motion resonances.
They applied $N$-body simulations and thermal emission calculations to
model this scenario and achieved a reasonable agreement with their
{\it IRAM} map.
An in-depth investigation on the Vega system dynamics was performed by
\citet{Wyatt-2003} who suggested that a Neptune-mass planet, migrating
outward from 40 to 65~AU over a time span of $\sim 56~\Myr$, may have
cleared the inner part of the assumed planetesimal disk and trapped a
significant amount of material in the $3:2$ and $2:1$ resonances, thus
creating two clumps as seen by \citet{Holland-et-al-1998}.
Later on, \citet{Reche-et-al-2008} generalized this theory to account for
eccentric planetary orbits.
Their findings are similar to those of \citet{Wyatt-2003} with the
difference that they require a Saturn-mass planet on a low eccentricity
orbit to account for the brightness asymmetry.
For the clumps to be visible against the non-resonant background, low
planetesimal eccentricities of $< 0.1$ are necessary.
Planets with masses greater than $\approx 2M_{\mathrm{Jupiter}}$ would
raise planetesimal eccentricities to about 0.2.
Besides, too massive planets would quickly deplete the disk.

As shown in section \ref{sec:excitation}, our simulations slightly favor
low planetesimal eccentricities up to 0.1.
This is in agreement with the limit given by \citet{Reche-et-al-2008}.
Still, a question arises whether non-inclusion of the azimuthal structure
in our simulations ({\it ACE} treats rotationally-symmetric disks) is a
reasonable assumption.
Indeed, \citet{Wyatt-2006} investigated the dust production in a clumpy
disk of resonant planetesimals and showed that local dust production from
the clumps is strongly enhanced and conversely, it is depressed between
the clumps.
We argue, however, that the net effect on the SED and radial profiles of
brightness is much weaker, because these depend on the
azimuthally-averaged dust production rates.
\citet{Queck-et-al-2007} found that the average collisional rate in a
resonant planetesimal belt is typically not more than twice as high as in
a similar non-resonant belt, while the average collisional velocities are
nearly unaffected by the resonant clumping.
Thus a collisional cascade in a resonant, clumpy belt can be approximated
by a cascade in a non-resonant, rotationally-symmetric belt with the same
mass at the same location, but somewhat higher orbital eccentricities of
planetesimals to mimic moderately enhanced collisional rates.
We stress that this is only valid when considering azimuthally-averaged
observables, not the azimuthally-resolved structure seen in the images.
For example, our approach is not suitable to make predictions for spiral
structure expected to emanate from the clumps.
\citet{Wyatt-2006} argues that such structure should be seen in mid- to
far-IR images, and that it is not, may simply be due to insufficient
resolution of the {\it Spitzer/MIPS} images or confusion in the
photospheric subtraction.

Throughout the study, we assumed that the initial eccentricities and
inclinations of the parent bodies in the planetesimal belt are
distributed according to energy equipartition.
This assumption would be reasonable if the distribution of orbits was
controlled by mutual collisions and gravitational scattering among
planetesimals, but it may not hold as soon as resonant interaction with
planets occurs.
A well-known example is our Edgeworth--Kuiper belt, in which the
eccentricities and inclinations of objects are distributed differently
\citep[e.g.][]{Brown-2001}.

Apart from the suspected planet that sculpts the main belt, the Vega
system may contain more  planets closer in.
In fact, a damped outward migration of the presumed planet that explains
the clumps {\em requires} the presence of another, more massive planet in
the system closer in \citep[e.g.][]{Gomes-et-al-2004}.
An inner planet, or planets, could stir the disk
\citep{Mustill-Wyatt-2009}.
Furthermore, several planets together could produce intricate combined
dynamical effects on the main planetesimal belt and its dust.
However, it seems premature to discuss them until new observations
have delivered evidence for these planets.

\subsection{The Exozodi in the Vega System}

As mentioned in the introduction, dust was surprisingly discovered in the
innermost part of the Vega system, inside $1$~AU
\citep{Absil-et-al-2006}.
Although reminiscent of the zodiacal cloud of the solar system, this
``exozodi'' of Vega remains a mystery. It seems to be far too dusty, and
the grain sizes retrieved from observations far too small, to be
explained by collisions in an ``asteroid belt'' or evaporation of comets.
One possibility would be a transport of planetesimals from the ``main''
debris disk inward and their subsequent disruption or evaporation.
Such a transport would require the presence of at least two planets.
In fact, a two-planet configuration~-- a ``Jupiter'' inside and a
``Saturn'' outside that shapes the main disk~-- could suffice
(Vandeportal et al., in prep.). Thus, the very existence of the exozodi
may strengthen the expectation that Vega hosts several planets, as
discussed above.

The direct contribution of the exozodi to the emission at $24\mum$
amounts to $\approx 0.6$~Jy.
This is about the emission which is lacking provided the used photometry
data is accurate.
However, as the exozodi could not be resolved with {\it Spitzer} and the
image is saturated at the stellar position, this very inner part of the
Vega disk cannot have affected the observation in the outer parts of the
system and can therefore be considered negligible.
Still, if not directly, the Vega exozodi could have an indirect impact on
the measured dust emission.
Dust inside 1~AU could have a shielding effect on dust located farther
out.
However, a simple estimate shows that the amount of stellar radiation to
which outer dust is exposed would only reduce by a factor of
$\sim 10^{-5}$.
Thus, we conclude that the very inner part of the system has no impact
on the outer disk's emission analyzed in this paper.

\section{CONCLUSIONS} \label{sec:summary}

Our analysis suggests that the debris disk of Vega is compatible with a
standard scenario, in which visible dust originates from a steady-state
collisional cascade operating in a ``Kuiper belt'', whose existence at
$\sim 100$~AU from the star is evident in sub-mm images.
We model the dust production from the sources, and find that thermal
emission of the resulting dust is fully consistent with the photometric
data across the entire wavelength range from mid-IR to radio covered by
observations.
Furthermore, we are able to naturally reproduce the radial brightness
profiles at $24$, $70$, and $160\mum$ derived from {\it Spitzer/MIPS}
observations.
Finally, the modeled emission agrees with the low-resolution images at
$850\mum$ taken with {\it SCUBA}.

If the Vega disk is maintained by a steady-state collisional cascade,
which appears likely, its total mass (in $\la 100~\km$-sized bodies) must
fall in the range from several to several tens of Earth masses.
Provided that collisional cascade has been operating over much of the
Vega age of $\sim 350~\Myr$, the disk must have lost a few Earth masses
of solids during that time.
Further constraints of the parameters of the system and physical
processes operating in the disk are as follows.
A reasonable amount of stirring should be present in the planetesimal 
ring.
We demonstrate that planetesimals are likely to have eccentricities of
the order of $\approx 0.1...0.3$, but the origin of stirring cannot be
constrained.
It may come, for instance, from a presumed giant planet interior to the
belt, or the disk can be self-stirred by largest, Pluto-sized
planetesimals.
Next, we show that the modeling results sensitively depend on the
luminosity of the central star.
Importantly, in the particular case of Vega, using a reduced radiation
flux from the stellar surface at low latitudes, which was derived from
its fast rotation, is mandatory to match modeled dust emission to the
data.
Another important prerequisite for this is to include cratering
collisions between the disk solids as a collisional outcome into the
model.

Another goal of this study was to use the Vega disk as a stringent test
for our modeling approach, which is a physical modeling of a debris disk
``from sources'' \citep{Krivov-et-al-2008}.
We deem the test successful.
We have shown that the approach does work and has a potential to deliver
constraints, most notably on the properties of directly invisible
planetesimals, that are not possible to put with other methods.
Our collisional and thermal emission model needs to be further tested and
``calibrated'' on other resolved debris disk systems.
Then, it can be used as a routine procedure in dynamical modeling of
debris disks, both known and expected to be detected with facilities like
{\it Herschel} or {\it ALMA}.

The data reduction and modeling presented here describe and explain the
outer system, outside $\approx 80~\AU$.
Inside that distance, our analysis of $24\mum$ emission observed by
{\it Spitzer/MIPS} indicates a possible rise of the $24\mum$ flux from
$10''$ inward, although it is not certain, because central part of the
{\it MIPS} $24\mum$ images is saturated.
Assuming, however, the data to be accurate, they cannot be explained by
the models presented here.
A natural explanation would be an additional dusty belt in the system at
$\sim 10~\AU$.
Such a belt could enhance the $24\mum$ emission coming from the ``main''
disk.
In the future, this hypothesis could be checked or falsified directly,
for instance with a more accurate mid-IR photometry of mid-IR
interferometry.
On any account, further observational and theoretical effort is required
to shed more light onto the inner part of the Vega system, including
alleged planets, possible planetesimal belts and dust rings, of which
one~--- an ``exozodi'' at just $\sim 1~\AU$~--- has recently been
discovered with near-IR interferometry.
It is the inner system that must eventually bring clues to the entire
architecture and the formation history of the Vega system.
A better knowledge of the inner part of the system, most notably
suspected planets there, will also result in a better understanding of
how the outer debris disk operates.

\begin{acknowledgements}
  We thank
  Olivier Absil,
  Jean-Charles Augereau,
  Herv{\'e} Beust,
  Hiroshi Kimura,
  Harald Mutschke,
  Kate Su,
  Phillippe Th\'ebault, and
  Mark Wyatt
  for numerous fruitful discussions on various aspects of this work.
  The review by the anonymous referee greatly helped to improve the paper.
  Part of this work was supported by the German
  \emph{Deut\-sche For\-schungs\-ge\-mein\-schaft, DFG\/} projects
  Kr~2164/5--1 and Kr~2164/8--1, by the \emph{Deutscher Akademischer
  Austauschdienst (DAAD)}, project D/0707543, and by the International
  Space Science Institute in Bern (``Exozodiacal Dust Disks and Darwin''
  working group\footnote{\sf http://www.issibern.ch/teams/exodust/}).
  SM is funded by the Graduate Student Fellowship of the Thuringia State.
  AK and TL are grateful to the Isaac Newton Institute for Mathematical
  Sciences in Cambridge where the final stages of this work were carried
  out in the framework of the program ``Dynamics of Discs and Planets''.
\end{acknowledgements}

\clearpage


\end{document}